\documentclass[aps,prb,superscriptaddress,twocolumn,longbibliography]{revtex4-2}

\usepackage{cancel}
\usepackage{physics}
\usepackage[english]{babel}
\usepackage[utf8]{inputenc}
\usepackage{amssymb}
\usepackage{amsmath}
\usepackage{amsfonts}
\usepackage{multirow}
\usepackage{float}
\usepackage[pdftex]{graphicx}
\usepackage{xspace}
\usepackage{dsfont}
\usepackage{soul}
\usepackage{graphics}
\usepackage{comment}
\usepackage{svg}

\usepackage{color,xcolor}

\usepackage{hyperref}
\hypersetup{colorlinks=true,citecolor=blue,linkcolor=blue,filecolor=blue,urlcolor=blue,pdfstartview=FitH,pdfpagemode=UseNone}
\usepackage{soul}
\usepackage[normalem]{ulem}
\definecolor{boxcolor}{HTML}{108f64}

\newcommand{\orcid}[1]{\href{https://orcid.org/#1}{\includegraphics[width=8pt]{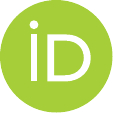}}}

\def\G{\Gamma}

\newcommand{\kvec}{\mathbf{k}}

\newcommand{\E}{\mathbf{E}}
\newcommand{\Eo}{\boldsymbol{\mathcal E}_0}

\begin{document}

\title{Surface lone-pair polarization probed by quantum-geometric transport in tellurium}

\author{Nathanael N. Batista \orcid{0000-0001-6103-3592}}
\affiliation{Departamento de F\'isica, Universidade Federal do Espírito Santo, 29075-910 Vitória, ES, Brazil}

\author{Wendel S. Paz \orcid{0000-0001-5737-0633}}
\affiliation{Departamento de F\'isica, Universidade Federal do Espírito Santo, 29075-910 Vitória, ES, Brazil}

\author{Manuel Suárez-Rodríguez \orcid{0000-0003-1186-2725}}
\affiliation{CIC nanoGUNE BRTA, 20018 Donostia-San Sebastián, Basque Country, Spain}

\author{Pierpaolo Fontana \orcid{0000-0002-8549-311X}}
\affiliation{Department of Physics, University of Trieste, Strada Costiera 11, I-34151 Trieste, Italy}

\author{Victor Velasco \orcid{0000-0003-3904-8602}}
\affiliation{International School for Advanced Studies (SISSA), Via Bonomea 265, I-34136 Trieste, Italy}

\author{Marcus V. O. Moutinho \orcid{0000-0003-0356-607X}}
\affiliation{Universidade Federal do Rio de Janeiro - Campus Duque de Caxias,
25240-005 Duque de Caxias, Brazil}

\author{Chang Niu \orcid{0000-0003-3175-7164}}
\affiliation{Elmore Family School of Electrical and Computer Engineering, Purdue University, West Lafayette, Indiana 47907, United States}
\affiliation{Birck Nanotechnology Center, Purdue University, West Lafayette, Indiana 47907, United States}

\author{Peide~D.~Ye \orcid{0000-0001-8466-9745}}
\affiliation{Elmore Family School of Electrical and Computer Engineering, Purdue University, West Lafayette, Indiana 47907, United States}
\affiliation{Birck Nanotechnology Center, Purdue University, West Lafayette, Indiana 47907, United States}

\author{Marco Gobbi~\orcid{0000-0002-4034-724X}}
\affiliation{IKERBASQUE, Basque Foundation for Science, 48009 Bilbao, Basque Country, Spain}
\affiliation{Centro de Física de Materiales CSIC-UPV/EHU, 20018 Donostia-San Sebastián, Basque Country, Spain}

\author{Fèlix Casanova \orcid{0000-0003-0316-2163}}
\affiliation{CIC nanoGUNE BRTA, 20018 Donostia-San Sebastián, Basque Country, Spain}
\affiliation{IKERBASQUE, Basque Foundation for Science, 48009 Bilbao, Basque Country, Spain}

\author{Luis E. Hueso \orcid{0000-0002-7918-8047}}
\affiliation{CIC nanoGUNE BRTA, 20018 Donostia-San Sebastián, Basque Country, Spain}
\affiliation{IKERBASQUE, Basque Foundation for Science, 48009 Bilbao, Basque Country, Spain}

\author{Caio Lewenkopf \orcid{0000-0002-2053-2798}}
\affiliation{Instituto de F\'isica, Universidade Federal do Rio de Janeiro, 21941-972 Rio de Janeiro, Brazil}

\author{Marcello B. Silva Neto \orcid{0000-0002-6817-3472}}
\affiliation{Instituto de F\'isica, Universidade Federal do Rio de Janeiro, 21941-972 Rio de Janeiro, Brazil}

\date{\today}

\begin{abstract}
Stereochemically active lone pairs are ubiquitous microscopic sources of polarity in molecules and solids, but their collective behavior in crystals is often hidden by symmetry or confined to surfaces. Here we show that quantum-geometry
transport provides a sensitive probe of surface lone-pair polarization in trigonal tellurium. 
This surface polarization appears microscopically as an inversion-odd dipolar component of the crystal potential, which shifts the center of mass of Bloch wavepackets and produces quantum-geometric
corrections to their velocity. 
We describe this lone-pair polar texture through a minimal three-component lattice model, and we show that the resulting linear and nonlinear transport coefficients probe, respectively, the second and first moments of the net polarization field. Because rectified voltages in tellurium flakes are directly proportional to the surface lone-pair polarization, our results provide a microscopic route to understanding and engineering polarization-driven, quantum-geometric electronic devices based on tellurium allotropes.
\end{abstract}

\maketitle

{\it Introduction.--}
Lone pairs of electrons are fundamental microscopic sources of polarity \cite{Gillespie1957,Payne2006,Payne2007}. 
In molecules, they determine geometry, enable directional bonding, and control the electric response through the stereochemical activity of nonbonding electronic states \cite{Gillespie1957}. In solids lone pairs can remain stereochemically active and strongly influence bonding and structural stability \cite{Cohen1992,laurita2022chemistry}. However, their collective thermodynamics is often difficult to access experimentally, as the polar distortions may be hidden in local or dynamic disorder, constrained by crystalline symmetry, or reshaped by surfaces and interfaces \cite{Fabini2020,laurita2022chemistry,Gattinoni2020,Wang2010}.

A paradigmatic example is trigonal tellurium \cite{Resurrection_2022}, 
consisting of helical chains whose interchain interactions are mediated by stereochemically active lone pairs \cite{NatureOfBonding2018}. 
Viewed in the basal plane, each unit cell hosts three symmetry-related in-plane lone-pair polar axes, whose directions are connected by the threefold rotational symmetry of the trigonal lattice \cite{NatureOfBonding2018,Dresselhaus2008}. 
In the bulk, this symmetry enforces an exact cancellation of the in-plane dipoles, leading to zero net in-plane polarization. 
In films or flakes, however, the loss of coordination at the surfaces breaks this $C_3$ equivalence, allowing 
an
in-plane polarization near the boundaries \cite{GiantNLHERectification}. 
Because both exposed surfaces select the same in-plane imbalance, their contributions add rather than cancel, making the surface polarization experimentally accessible when the corresponding bulk response is forbidden by symmetry \cite{Sodemann2015,Du2021,GiantNLHERectification}.

\begin{figure}[!t]
\centering
\includegraphics[width=0.95\linewidth]{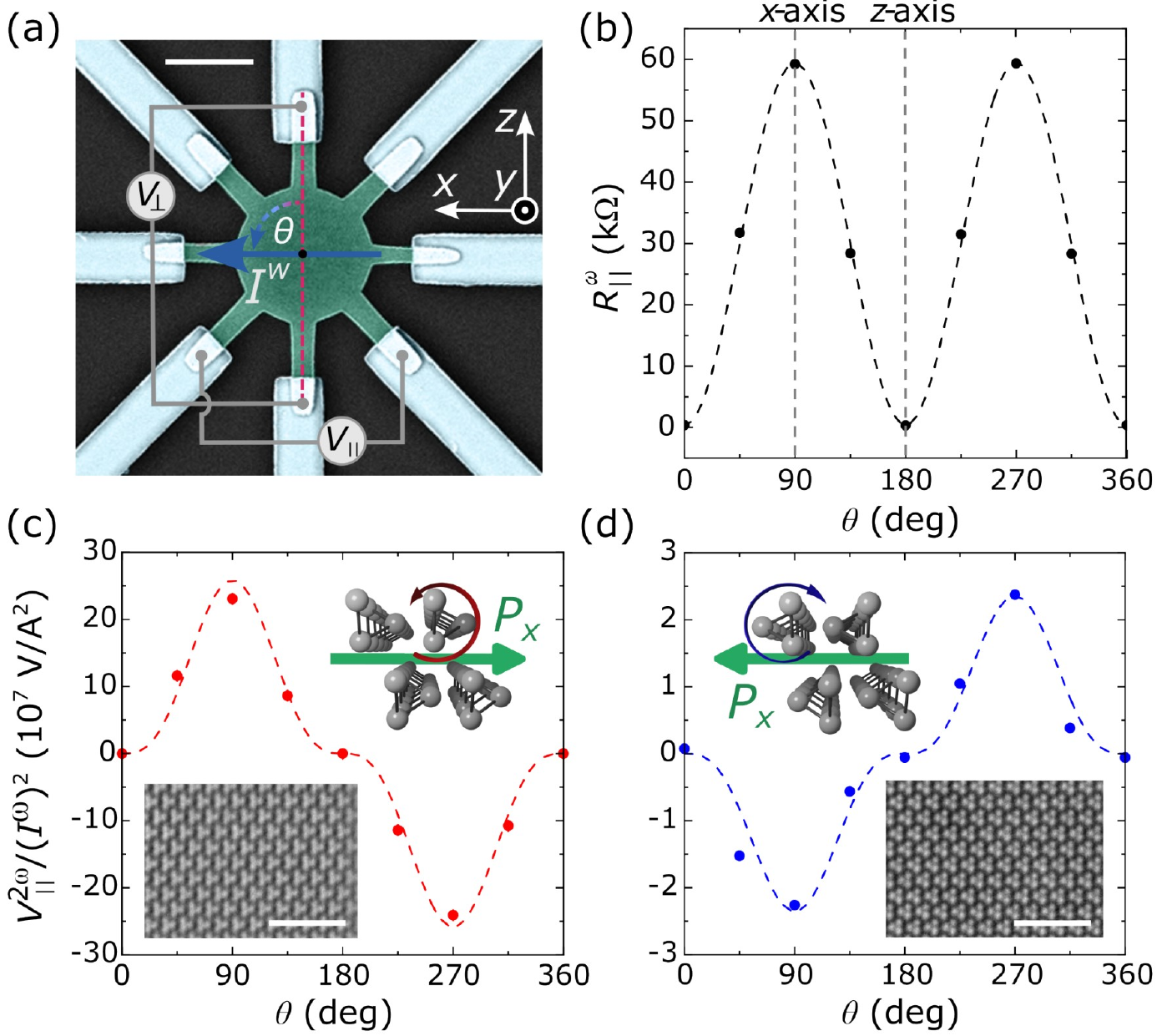}
\caption{
Quantum-geometric transport.
Experimental data from Ref.~\cite{SuarezRodriguez2024}. (a) An AC current $I^\omega$ is injected while the sample is rotated in the $xz$ plane by an angle $\theta$ with respect to the crystallographic axes. (b) First-harmonic longitudinal resistance $V_\parallel^\omega/I^\omega=R_\parallel^\omega$ measured as a function of $\theta$. The linear response is proportional to $-\delta\sigma^{geo}_{xx}({\cal E}_0^2)\,\sin^2\theta$, with $\delta\sigma^{geo}_{xx}<0$ for any ${\cal E}_0$. (c) and (d) Second-harmonic $V_\parallel^{2\omega}/(I^\omega)^2$ also measured longitudinally under the same conditions. The nonlinear response is proportional to $-\chi_{xxx}({\cal E}_0)\,\sin^3\theta$, with (c) $\chi_{xxx}<0$ for ${\cal E}_0>0$ and (d) $\chi_{xxx}>0$ for ${\cal E}_0<0$. Inset: STEM images from Ref.~\cite{SuarezRodriguez2024} showing the trigonal structure for samples of opposite chirality and related by an inversion operation; ${\cal E}_0$ represents the lone-pair polarization texture shown as $P_x$ (green arrow) with ${\cal E}_0>0$ for (c) and ${\cal E}_0<0$ for (d).
}
\label{fig:device_angular_1st_and_2nd}
\end{figure}

Recent transport experiments in tellurium flakes reported a giant anisotropy in the 
linear resistivity 
$\rho_{xx}$ and a large longitudinal second-harmonic response, $\chi_{xxx}$
at zero magnetic field, featuring a characteristic in-plane angular dependence of the applied current with respect to the crystallographic direction, see Fig.~\ref{fig:device_angular_1st_and_2nd}(a), as well as a strong dependence on 
gating \cite{SuarezRodriguez2024}. 
Within the standard framework of intrinsic nonlinear transport in time-reversal-invariant metals, the Berry-curvature dipole is often identified as the dominant mechanism and is typically associated with Hall-like responses \cite{Sodemann2015,Ortix2021,suarez2025nonlineartransport}. 
The observed dominance of $\rho_{xx}$ and $\chi_{xxx}$, together with their unique gate-voltage and temperature dependences, therefore raises a fundamental question: what microscopic degree of freedom is being probed by these linear and nonlinear transport signals?

In this work, we demonstrate that a net surface lone-pair polarization texture in tellurium, represented by $\Eo$, produces a quantum-geometry
correction to the semiclassical wavepacket velocity whose leading inversion-odd contribution is proportional to $\Eo$ \cite{Gravity,Kaplan_2024_PRL,Sundaram1999,Gao2014}. 
Within Boltzmann transport theory, this term generates a dominant longitudinal nonlinear conductivity $\chi_{xxx}<0$ (for ${\cal E}_0>0$), that determines the second-harmonic response $V_\parallel^{2\omega}/(I^\omega)^2\propto -\chi_{xxx}/\sigma_D^3$ \cite{SuarezRodriguez2024}, where $\sigma_D$ is the Drude conductivity and $I^\omega$ is the driving current. 
Concurrently, $\Eo$ produces an inversion-even geometric correction $\propto \Eo^2$ to the linear conductivity, $\delta\sigma^{geo}_{xx}<0$, which governs the first-harmonic response $V^{\omega}/I^\omega\propto -\delta\sigma^{geo}_{xx}/\sigma_D^2$ \cite{SuarezRodriguez2024}. 
By comparing their temperature dependences with experiments, using a minimal lattice model, we demonstrate that $\chi_{xxx}$ and $\delta\sigma^{geo}_{xx}$ probe the first $\langle\Eo\rangle_T$ and second  $\langle\Eo^2\rangle_T$ moments of the lone-pair polarization texture, respectively. 
Finally, since $\chi_{xxx}$ controls the DC rectification in tellurium ($V_{\rm DC}\propto\chi_{xxx}\propto\langle\Eo\rangle_T$) \cite{Suarez-Adv-Mat-Rectification}, our work provides a microscopic route to engineering  surface lone-pair polarization,
paving the way for the design of controllable, radio-frequency quantum-geometric (QG) rectifiers based on tellurium nanostructures.

\begin{figure}[t]
\centering
\includegraphics[width=0.85\linewidth]{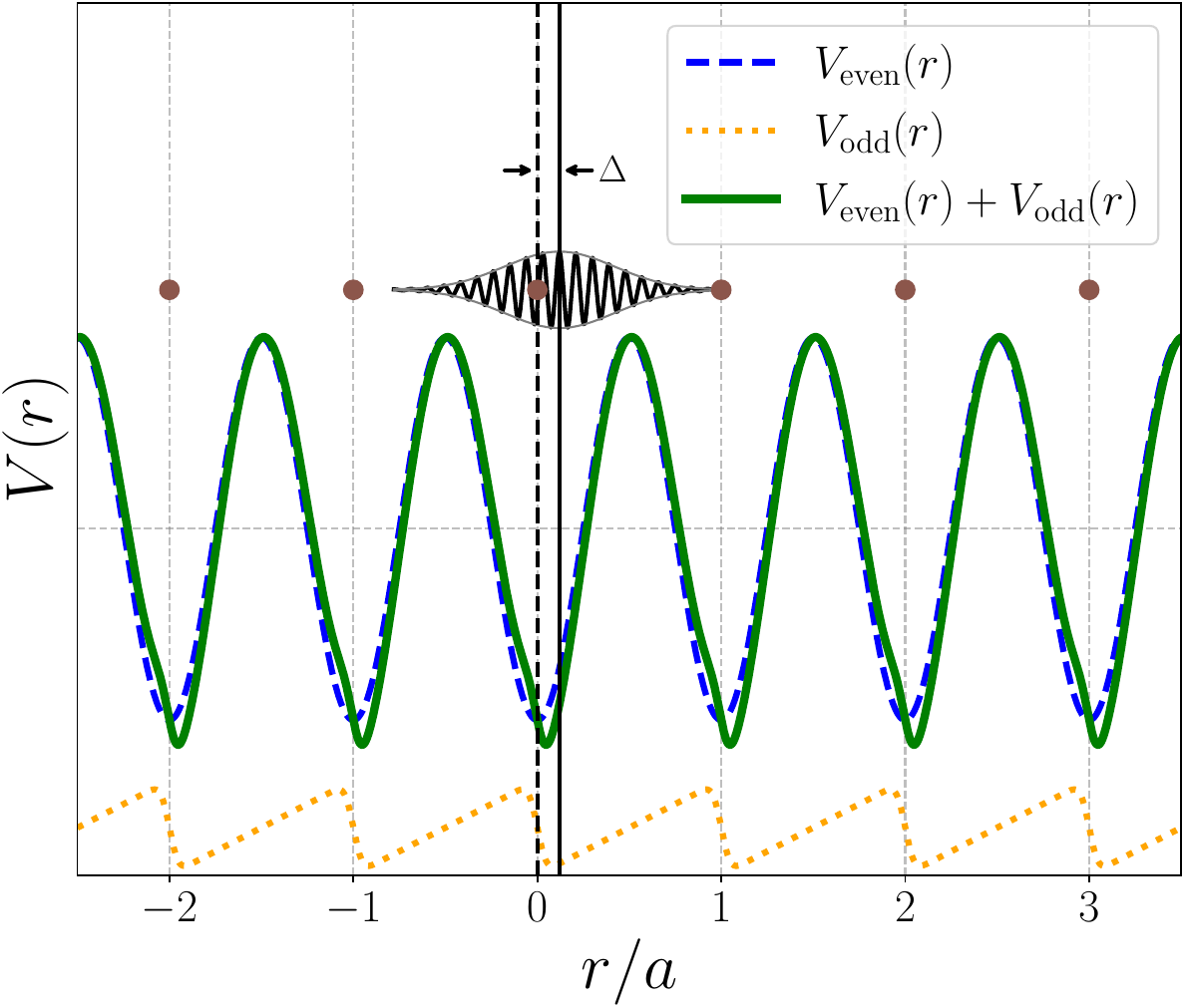}
\caption{
Inversion-odd wavepacket shift. 
Schematic illustration of the one-dimensional crystal potential, $V(r) = V_{\rm even}(r) + V_{\rm odd}(r)$. 
The inversion-even component $V_{\rm even}(r)$ defines symmetric wells centered at the ionic positions $r_0$ (brown circles). 
The inversion-odd component $V_{\rm odd}(r)$ introduces an asymmetric local slope $-e\mathcal{E}_0 \equiv \nabla V_{\rm odd}|_{r_c \simeq r_0}$, which shifts the total potential minimum away from the centrosymmetric reference (dashed line) by an offset $\Delta$ (solid line). 
This inversion-breaking crystal field causes a positional shift in the center of mass of the localized Bloch wavepacket.
}
\label{fig:polar_dipolar_shift}
\end{figure}

{\it Inversion-odd wavepacket dynamics.--} 
The surface lone-pair polarization enters the microscopic Hamiltonian as an inversion-odd polar component of the crystal potential, $V(\mathbf r)$, producing interband dipolar mixing and a positional shift of the wavepacket center of mass. 
We therefore start with a multi-band, ${\bf k}\cdot{\bf p}$ Hamiltonian for the valence-band (VB) structure near the $H$ point \cite{Nakao_Doi_Kamimura_I}
\begin{equation}
\hat H(\mathbf k)=\frac{\hat{\mathbf p}^2}{2m}+
V(\mathbf r)+\frac{\hbar}{m}\mathbf k\cdot\hat{\mathbf p}+
\frac{\hbar}{4m^2c^2}\left[\nabla V(\mathbf r)\times \hat{\mathbf p}\right]\cdot\boldsymbol\sigma^{(s)},
\label{eq:microscopic_kp_main}
\end{equation}
where $m$ is the electron mass, $c$ is the speed of light, and $\sigma^{(s)}$ are spin Pauli matrices. 
The crystal potential is decomposed as $V(\mathbf r)=V_{\rm even}(\mathbf r)+V_{\rm odd}(\mathbf r)$, where the inversion-odd component $V_{\rm odd}(\mathbf r)$ models the lone-pair polar field $\boldsymbol{\mathcal{E}}_0$, and we neglect $k$-dependent spin-orbit terms. 
Next, we project Eq.~\eqref{eq:microscopic_kp_main} onto the two uppermost valence bands, $H_4$ and $H_5$ \cite{iacovelli2026polar}. 
Denoting the projector onto this subspace as $P_{45}$ and its complement as $Q=1-P_{45}$, the effective VB Hamiltonian is obtained via L\"owdin downfolding,
\begin{equation}
\hat H^{\rm eff}_{45}(\mathbf k)=P_{45}\hat H(\mathbf k)P_{45}+
P_{45}\hat H(\mathbf k)Q\frac{1}{\Delta E_{P-Q}}
Q\hat H(\mathbf k)P_{45},
\label{eq:lowdin_main}
\end{equation}
where $\Delta E_{P-Q}=\varepsilon_{45}-Q\hat H(\mathbf k)Q$, with $\varepsilon_{45}$ denoting the unperturbed energy of the doublet.
The detailed derivation of this projection is given in the Supplemental Material \cite{SM, cardy, Ozaki2003, OzakiKino2004, OzakiKino2005, OzakiTotalEnergy, OzakiBerryPhase, pbe, monkhorst1976special,KingSmithVanderbilt1993, Resta1994}. 
In this projected language, the local gradient of $V_{\rm odd}$ manifests as a projected dipolar matrix element $P_{45}\left[\nabla V_{\rm odd}\cdot(\hat{\mathbf r}-\mathbf r_c)\right]P_{45}$, defined in terms of the wavepacket center-of-mass coordinate $\mathbf r_c$ and the internal coordinate $\hat{\mathbf r}$. 
This allows us to express the polarization contribution within the projected subspace as $\hat H_{\rm pol}=-e\,\boldsymbol{\mathcal{E}}_0\cdot(\hat{\mathbf r}-\mathbf r_c)$, where $\left.\nabla V_{\rm odd}\right|_{\mathbf r_c\simeq\mathbf r_0}\equiv -e\,\boldsymbol{\mathcal{E}}_0$ represents the local slope of the inversion-odd crystal potential at the ionic position $\mathbf r_0$ (see Fig.~\ref{fig:polar_dipolar_shift}).
For a crystal with more than one atom in the unit cell, this definition generalizes to each atomic position; the polar scale entering transport is its corresponding unit-cell average, as discussed below.

The coupling to an applied electric field has the usual electric-dipole form, $\hat H_E=-e\,\mathbf E\cdot(\hat{\mathbf r}-\mathbf r_c)$ \cite{Sundaram1999,XiaoRMP}. 
Since the inversion-odd polar crystal potential $\hat H_{\rm pol}$ has the same operator structure as $\hat H_E$, the total dipolar operator responsible for interband mixing is simply
\begin{equation}
    \hat W_{\rm dip}=-e[\mathbf E+\Eo]\cdot(\hat{\mathbf r}-\mathbf r_c).
    \label{eq:Wdip}
\end{equation}
An electric-dipole perturbation such as Eq.~\eqref{eq:Wdip} admixes remote bands into the wavepacket and produces a gauge-invariant positional shift of its center of mass, denoted by $\delta r_{c,i}\equiv a'_{n,i} =eG^{(n)}_{ij}(\mathbf k)[E_j+{\mathcal{E}}_{0,j}]$, where $G^{(n)}_{ij}$ is a cross-gap tensor built from interband matrix elements \cite{Kaplan_2024_PRL,Gao2014}. 

The same interband reconstruction also yields a second-order quantum-geometric correction to the wavepacket energy
\begin{equation}
    \delta\tilde{\varepsilon}^{\rm geo}_n(\mathbf k)=
    e^2\Lambda^{(n)}_{ij}(\mathbf k)
    [E_i+{\mathcal{E}}_{0,i}][E_j+{\mathcal{E}}_{0,j}],
    \label{eq:geo_energy}
\end{equation}
where $\Lambda^{(n)}_{ij}(\mathbf k)$ is the energy-normalized Berry connection susceptibility \cite{Gao2014,Kaplan_2024_PRL}. 
While distinct from the standard intrinsic quantum metric \cite{suarez2025nonlineartransport}, its numerator is constructed from the same real interband Berry-connection matrix elements, whereas its energy denominator accounts for the spectral cost of virtual transitions to remote bands \cite{Gao2014,Kaplan_2024_PRL}.
Thus, $\Lambda^{(n)}_{ij}(\mathbf k)$ is a quantum-metric-related response tensor whose derivative $\Gamma^{(n)}_{\alpha ij}(\mathbf k)\equiv\partial_{k_\alpha}\Lambda^{(n)}_{ij}(\mathbf k)$ defines what we refer to as the QG correction to the velocity, $v^{\rm geo}_{n,\alpha}(\mathbf k)=(1/\hbar)\partial_{k_\alpha} \delta\tilde{\varepsilon}^{\rm geo}_n(\mathbf k)$,
\begin{equation}
    v^{\rm geo}_{n,\alpha}(\mathbf k)=
    \frac{e^2}{\hbar}
    \Gamma^{(n)}_{\alpha ij}(\mathbf k)
    [E_i+{\mathcal{E}}_{0,i}][E_j+{\mathcal{E}}_{0,j}].
    \label{eq:vgeo_general_main}
\end{equation}
%

{\it Quantum-geometric transport.--}
The calculation of conductivities starts with the definition of current \cite{Ziman}
\begin{equation}
    \mathbf j(t)=-e\sum_n\int_{\mathbf k}\mathbf v_n(\mathbf k,t)\,f_n(\mathbf k,t),
    \quad\int_{\mathbf k}\equiv\int\frac{d^3k}{(2\pi)^3}.
    \label{eq:current_definition_main}
\end{equation}
Within the relaxation-time approximation, $\tau_{\mathbf k}=\tau$, and in the low-frequency regime $\omega\tau\ll1$, the linear correction to the carrier distribution function induced by the field-driven acceleration $\dot{\mathbf k}=-(e/\hbar)\mathbf E$ is  \cite{Ziman}
\begin{equation}
    f_n^{(1)}(\mathbf k,t)\simeq\frac{e\tau}{\hbar}
    \mathbf E\cdot \nabla_{\mathbf k} f_n^{(0)},
    \label{eq:f1_new}
\end{equation}
where $\nabla_{\mathbf k}f_n^{(0)}=\hbar \mathbf v_{n}(\mathbf k)\partial_\varepsilon f^{(0)}$.

\begin{figure}[t]
\centering
\includegraphics[scale=0.50]{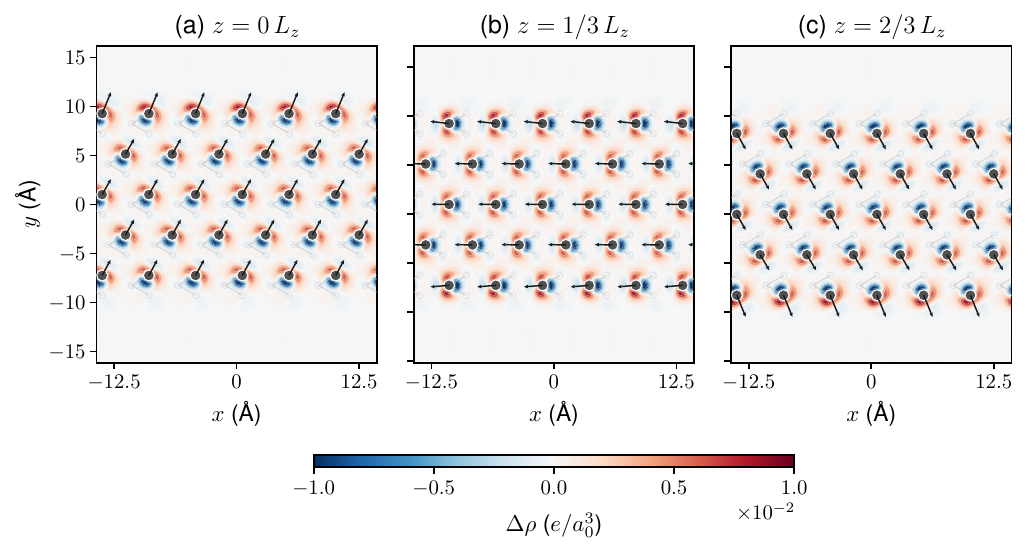}
\caption{
Excess electronic density (red) and charge depletion (blue) relative to the atomic densities of constituent atoms.
The three panels correspond to slices at (a) $z = 0 $, (b) $z = L_z/3$, and (c) $z = 2L_z/3$, for the three non-equivalent atoms in the unit cell. 
The black arrows indicate the local inversion-odd dipolar slopes computed from the gradient of the Hartree potential, providing the directions of the local lone-pair dipoles. 
}
\label{fig:Crystal_Field_ELF_Tellurene}
\end{figure}

Trigonal tellurium contains three inequivalent Te-atoms in the unit cell \cite{Resurrection_2022,NatureOfBonding2018}. 
Therefore, the dipolar scale in
the long-wavelength transport theory, Eq.~\eqref{eq:vgeo_general_main}, is the unit-cell average of the three local polar slopes
\begin{equation}
    \Eo=\frac{1}{3}\sum_{a=1}^{3}{\Eo}_a,
    \qquad{\Eo}_a\equiv -\frac{1}{e}\nabla V_{\rm odd}\Big|_{\mathbf r_c\simeq\mathbf r_a},
    \label{eq:E0_unit_cell_average}
\end{equation}
where $\mathbf r_a$ labels the three Te positions inside the unit cell. Our DFT calculations, shown in Fig.~\ref{fig:Crystal_Field_ELF_Tellurene},
provide a direct microscopic visualization of these local polar slopes. 
In the bulk, the three local gradients of the Hartree potential associated with the three 
Te sites are related by $C_3$ structure and cancel in the unit-cell average, so that $\Eo=0$, despite the presence of strong local dipolar textures. 
Near the two surfaces, however, the loss of interchain coordination  breaks the equivalence of the three local environments promoting a surface redistribution of charge. 
The cancellation is then incomplete, producing a finite  surface contribution $\Eo\parallel\hat{\mathbf x}$, in full agreement with earlier experimental observations \cite{Chang_2023, GiantNLHERectification, FontanaPRL2025}. 
In what follows, we therefore parametrize the external drive relative to the special $\hat{\mathbf x}$ axis as $\E(t)=E^{\omega}\sin\theta\cos\omega t\,\hat{\mathbf x}$, 
where $\theta$ denotes the angle relative to the polarization direction.

{\it First-harmonic response.--}
The first-harmonic current contains the usual Drude contribution plus the geometric correction proportional to $\boldsymbol{\mathcal E}_0^2$ from Eq.~\eqref{eq:vgeo_general_main}, namely
\begin{equation}
    j_x^\omega(t)=-e\sum_n\int_{\mathbf k}
    \left[v_{x,n}(\mathbf k)+\frac{e^2}{\hbar}\Gamma^{(n)}_{xxx}(\mathbf k)\mathcal E_0^2\right]
    f_n^{(1)}(\mathbf k,t).
    \label{eq:jw_first_harmonic_main}
\end{equation}
From Ohm's law, $j_x^\omega(t)=\sigma_{xx}E^{\omega}\sin\theta\cos\omega t$, the conductivity along the polar axis is $\sigma_{xx}=\sigma_{xx}^{D}+ \delta\sigma_{xx}^{\rm geo}$.
Here, the Drude term is $\sigma_{xx}^{D}=e^2\tau\sum_n\int_{\mathbf k}v_{x,n}^2(\mathbf k)(-\partial_\varepsilon f^{(0)})$ and
\begin{equation}
    \delta\sigma_{xx}^{\rm geo}=
    \frac{e^4\tau\,\mathcal E_0^2}{\hbar}
    \sum_n\int_{\mathbf k}\Gamma^{(n)}_{xxx}(\mathbf k)v_{x,n}(\mathbf k)
    \left(-\partial_\varepsilon f^{(0)}\right).
    \label{eq:sigma_geo}
\end{equation}
Equation~\eqref{eq:sigma_geo} shows that the linear longitudinal correction is inversion-even ($\propto\mathcal E_0^2$) and is nonzero because because the Fermi-surface average $\langle\Gamma^{(n)}_{xxx}(\mathbf k)v_{x,n}(\mathbf k)\rangle_{FS}\neq0$. 
The resulting conductivity tensor is diagonal, with principal components $\sigma_{xx}$ and $\sigma_{zz}$. 
Inverting 
this tensor gives the resistivity $\rho(\theta)=\rho_{zz}\cos^2\theta+
    \rho_{xx}\sin^2\theta,
    \label{eq:rho_theta}$
with $\rho_{ii}=1/\sigma_{ii}$, which precisely reproduces the angular dependence observed experimentally in the first-harmonic resistance shown in Fig.~\ref{fig:device_angular_1st_and_2nd}(b).

\begin{figure}[t]
\includegraphics[width=\linewidth]{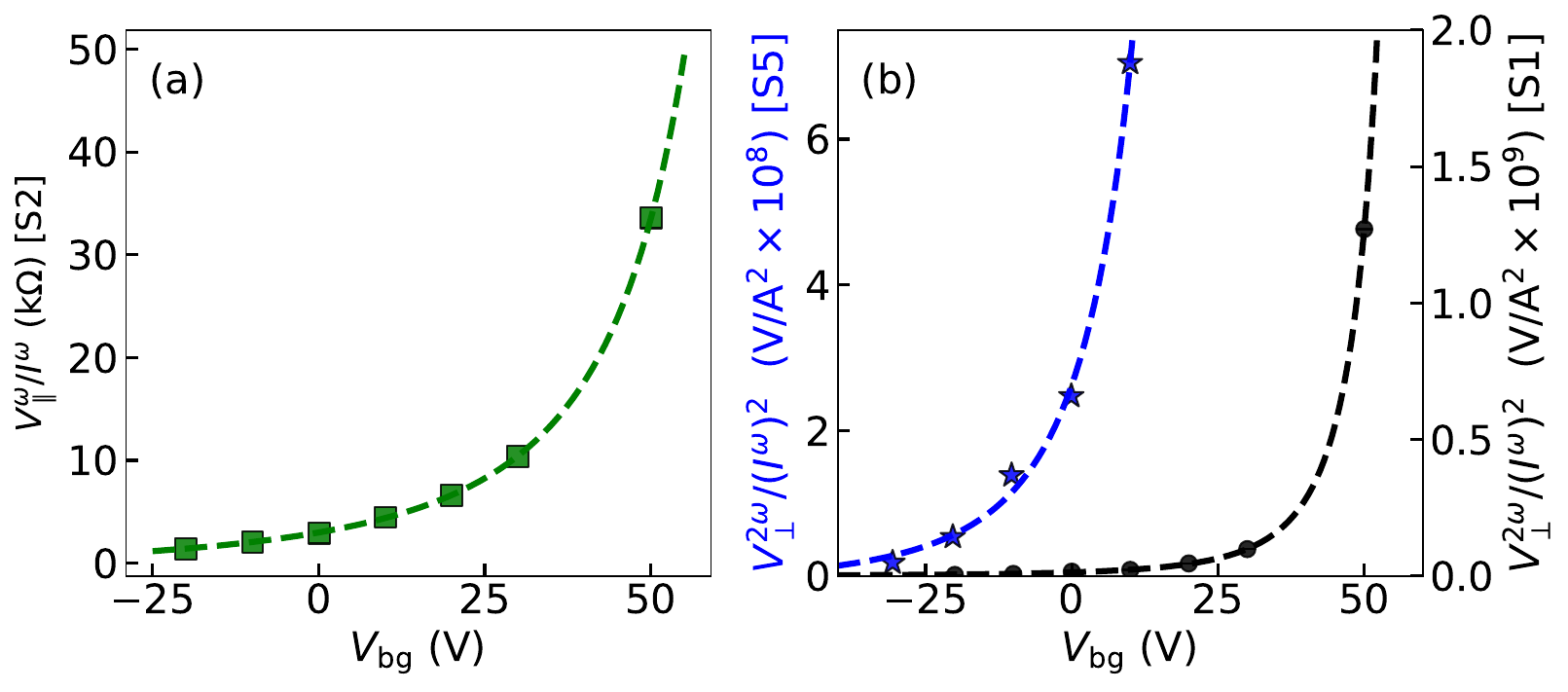}
\caption{
Gate-voltage dependence. 
Experimental data from Ref.~\cite{suarez2025Thesis} showing scaling behavior of the harmonic responses as a function of the applied back-gate voltage $V_{bg}$ for selected samples (labeled on the $y$-axis). 
(a) First-harmonic response, $V_\parallel^\omega/I^\omega \sim \mu^{-5/2}$, and (b) second-harmonic response, $V_\perp^{2\omega}/(I^\omega)^2 \sim \mu^{-7/2}$, displaying the scaling laws predicted from the QG correction to $\delta\sigma^{\rm geo}_{xx}$ and  $\chi_{xxx}$, respectively.
}
\label{fig:Gate-Dependence-First-Second-Harmonics}
\end{figure}

{\it Second-harmonic response.--}
The nonlinear longitudinal current, $j_x^{2\omega}$, arises from the component of the geometric 
velocity proportional to $2\mathcal E_0E^{\omega}_x(t)$ in Eq.~\eqref{eq:vgeo_general_main} 
combined with the Boltzmann correction $f^{(1)}$ from Eq. \eqref{eq:f1_new}:
\begin{align}
    \nonumber
    j_x^{2\omega}=&
    \frac{e^4\tau\,\mathcal E_0}{\hbar}
    (E^{\omega})^2\sin^2\theta
    \sum_n\int_{\kvec}\!
    \G^{(n)}_{xxx}(\kvec)v_{x,n}(\kvec)
    \left(\!-\partial_\varepsilon f^{(0)}\!\right).
\end{align}
Defining the nonlinear current as $j_x^{(2)}(t)=\chi_{xxx}^{\rm geo}E_x^2(t)$, the $2\omega$ amplitude is
given by $j_x^{2\omega}=\chi_{xxx}^{\rm geo}(E_x^\omega)^2/2$, yielding the nonlinear quantum-geometric conductivity
\begin{equation}
    \chi_{xxx}^{\rm geo}=
    \frac{2e^4\tau\,\mathcal E_0}{\hbar}
    \sum_n\int_{\kvec}
    \G^{(n)}_{xxx}(\kvec)v_{x,n}(\kvec)
    \left(-\partial_\varepsilon f^{(0)}\right)\neq 0.
    \label{eq:chi_final_new}
\end{equation}
Projecting the electric field onto the basis
$\hat{\mathbf e}_\parallel=(\sin\theta,0,\cos\theta)$ and
$\hat{\mathbf e}_\perp=(\cos\theta,0,-\sin\theta)$ gives \cite{suarez2025symmetryorigin}
\begin{eqnarray}
    E^{2\omega}_\parallel &\propto& \sin\theta\,E_x^{2\omega}
    \propto \sin^3\theta,
    \label{eq:sin-cube}\\
    E^{2\omega}_\perp &\propto&
    \cos\theta\,E_x^{2\omega}
    \propto \sin^2\theta\cos\theta .
\end{eqnarray}
%
The angular dependence predicted by Eq.~\eqref{eq:sin-cube} is precisely the one observed in the longitudinal second-harmonic voltages shown in Figs.~\ref{fig:device_angular_1st_and_2nd}(c) and (d) for different samples related by an inversion operation.

Finally, the quantum-geometric framework also predicts the absence of a nonlinear longitudinal response for currents flowing along the helical chains, that is, at $\theta=0$ ($\hat{\mathbf z}$ direction), consistent with group theoretical analysis \cite{suarez2025symmetryorigin}. 
Here, since $\Eo\parallel\hat{\mathbf x}$, the relevant Christoffel symbol is
$\Gamma^{(n)}_{zxz}$. 
However, $v_z(\kvec)$ is odd in $k_z$, while $\Gamma_{zxz}(\kvec)$ is even in $k_z$ and odd in $k_x$ \cite{SM}. 
Their product is therefore odd in momentum, causing its Fermi-surface average to vanish identically, $\langle\Gamma^{(n)}_{zxz}(\mathbf k)v_{z,n}(\mathbf k)\rangle_{FS}\equiv 0$. 
Consequently,
\begin{equation}
    \chi^{geo}_{zzz}\propto\sum_n\!\int_{\kvec} v_{z,n}(\kvec)\Gamma^{(n)}_{zxz}(\kvec)\left(-\partial_\varepsilon f^{(0)}\right)\equiv 0.
\end{equation}

{\it Gate-voltage dependence.--} 
To demonstrate that $\delta\sigma_{xx}^{geo}$ and $\chi_{xxx}$ are of QG origin we start from the relations 
\begin{equation}
    \frac{E_\parallel^{\omega}}{j_x^\omega}=\rho_D-\frac{\delta\sigma_{xx}^{geo}}{\sigma_D^2}\sin^2\theta,\quad
    \frac{E_\perp^{2\omega}}{(j_x^\omega)^2}=-\frac{1}{2}\frac{\chi_{xxx}}{\sigma_D^3}\sin^2\theta\cos\theta.
    \label{eq:V2w_scaling}
\end{equation}
From Eqs.~\eqref{eq:sigma_geo} and \eqref{eq:chi_final_new} we find that the chemical potential dependence of both $\delta\sigma_{xx}^{geo}$ and $\chi_{xxx}$ at low temperature is determined by $\sum_n\langle \G^{(n)}_{xxx}(\kvec)v_{x,n}(\kvec)\rangle_{FS}$. 
For the VB of tellurene we have \cite{SM} $\langle \G^{(n)}_{xxx}(\kvec)v_{x,n}\rangle_{FS}\sim \mu^{-1/2},\quad\mbox{and}\quad \sigma_D\sim\mu$, which gives the QG 
scaling for the measured responses $V_\parallel^{\omega}/I^\omega-\rho_D\propto \mu^{-5/2}$ and $V_\perp^{2\omega}/(I^\omega)^2\propto \mu^{-7/2}$ as a function of back-gate voltage. 
The close agreement between this theory and 
 the experimental gating characteristics shown in Figs.~\ref{fig:Gate-Dependence-First-Second-Harmonics}(a) at $\theta=90^\circ$ and (b) at $\theta=45^\circ$ strongly supports the QG
origin of both $\delta\sigma_{xx}^{\rm geo}$ and $\chi_{xxx}$. 

{\it Model for surface polarization.--}
We now turn to the microscopic model of the polar scale $\Eo$ in tellurium. 
We take the three local polar directions entering Eq.~\eqref{eq:E0_unit_cell_average} as
\begin{equation}
    \hat {\bf u}_1=(-1,0),\;\;
    \hat {\bf u}_2=\left(\tfrac12,-\tfrac{\sqrt3}{2}\right),\;\;
    \hat {\bf u}_3=\left(\tfrac12,+\tfrac{\sqrt3}{2}\right),
\end{equation}
which satisfy $\hat {\bf u}_1+\hat {\bf u}_2+\hat {\bf u}_3=0$. 
This constraint introduces a geometric frustration at the local level: 
although the local dipoles at each triangular motif $t$ develop a finite magnitude, $w_{t,a}$, the unit-cell polarization ${\bf P}_t=p_0\sum_{a=1}^3w_{t,i}\,\hat {\bf u}_a$ vanishes in the bulk, when all three local slopes are equivalent, $w_{t,1}=w_{t,2}=w_{t,3}$. 

\begin{figure}[t]
\centering
\includegraphics[width=\linewidth]{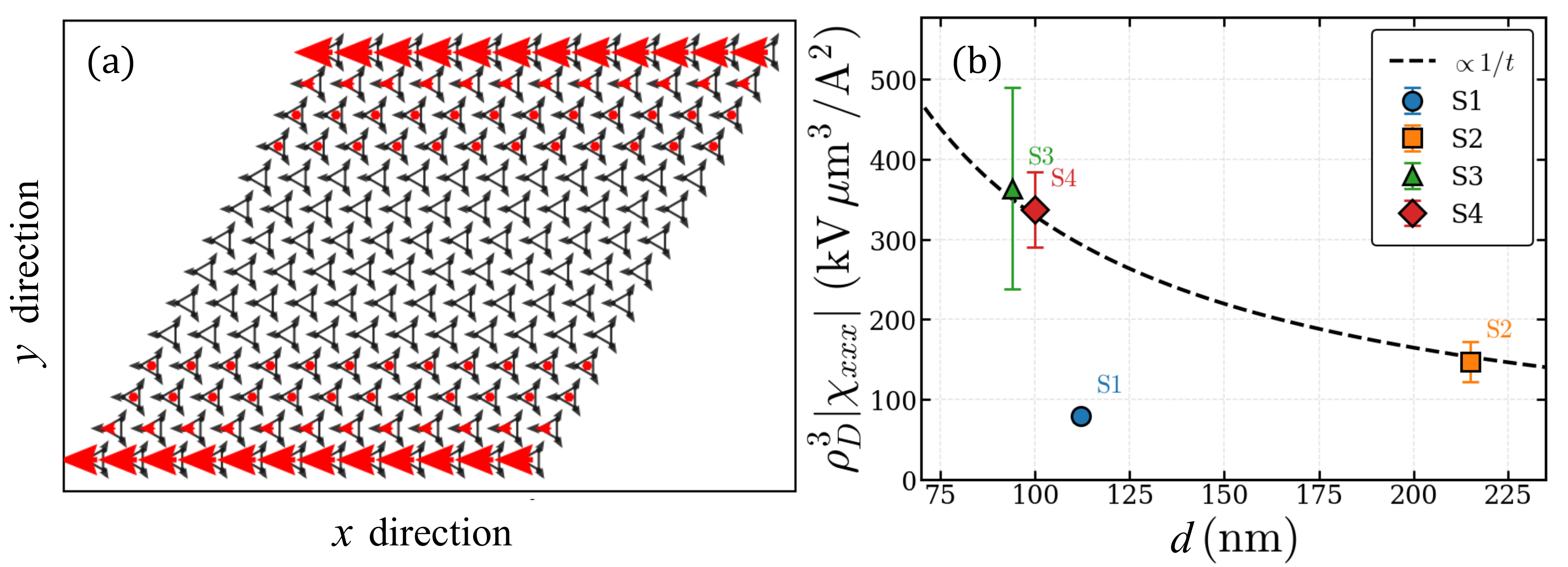}
\caption{
{Lone-pair polarization.}
(a) Schematic showing how the loss of coordination at the boundaries breaks the bulk $C_3$ symmetry, generating a finite polarization near the surfaces. 
Here, $\Eo<0$ (and thus $\chi_{xxx}(\Eo)>0$) produce the angular dependence shown in Fig.~\ref{fig:device_angular_1st_and_2nd}(d). 
(b) Measured $\rho_D^3|\chi_{xxx}(d)|$ as a function of flake thickness $d$ for samples S1, S2, S3, and S4 \cite{suarez2025Thesis}, demonstrating a clear $1/d$ dependence.
The deviation of S1 from this scaling is discussed in the main text.}
\label{fig:Polarization_bulk}
\end{figure}

However, in a film cut along the $\hat{{\bf y}}$ direction, tellurium atoms at the top and bottom surfaces lose coordination with neighboring chains located above and below the film [see Figs. \ref{fig:Crystal_Field_ELF_Tellurene}(a)-(c)]. 
A coarse-grained energy in terms of the polarization amplitudes $w_{t,a}$ can be written as
\begin{align}
    \nonumber
    E=&\frac{\kappa}{2}\sum_{t,a}(w_{t,a}-1)^2+
 \frac{J}{2}\sum_{\langle t,t' \rangle,a}(w_{t,a}-w_{t',a})^2\\
    &-\sum_t h(j_t)\,w_{t,1}.
    \label{eq:energy_triangles}
\end{align}
Here $\kappa$ enforces restoration towards uniform dipole weights, $J$ penalizes spatial variations via inter-triangle coupling between adjacent layers $\langle t,t' \rangle$, and $h(j)$ is a surface-induced symmetry-breaking field, $h(j)=h_0 e^{-d(j)/\lambda_s}$, where $d(j)$ denotes the distance of layer $j$ from the nearest surface and $\lambda_s$ is a screening length that dictates how far the boundary-induced bias $h(j)$ extends into the bulk. 
The competition between $J$ and $\kappa$ yields an exponential decay of the induced polarization, $P_x(j)\sim e^{-j/\xi}$, with $\xi\sim \sqrt{J/\kappa}$, shown as surface red arrows in Fig.~\ref{fig:Polarization_bulk}(a), leaving a completely non-polar bulk.

The average polarization scales as $\langle P_x\rangle\sim 2\xi/N_y$ as a function of the number of layers $N_y$, as validated by our DFT calculations \cite{SM}. 
Experimentally, the measured $\rho_D^{3}|\chi_{xxx}(d)|$ [see Fig.~\ref{fig:Polarization_bulk}(b)], which is directly proportional to the average polarization, $\langle P_x\rangle(d)$, follows a $1/d$ scaling law as a function of the flake thickness $d$ for samples S2, S3, and S4 \cite{suarez2025symmetryorigin}. 
The sole outlier is sample S1, which was removed from the cryostat in between measurements, causing variations in doping level, thereby reducing the measured signal. 

\begin{figure}[!t]
\centering
\includegraphics[width=\linewidth]{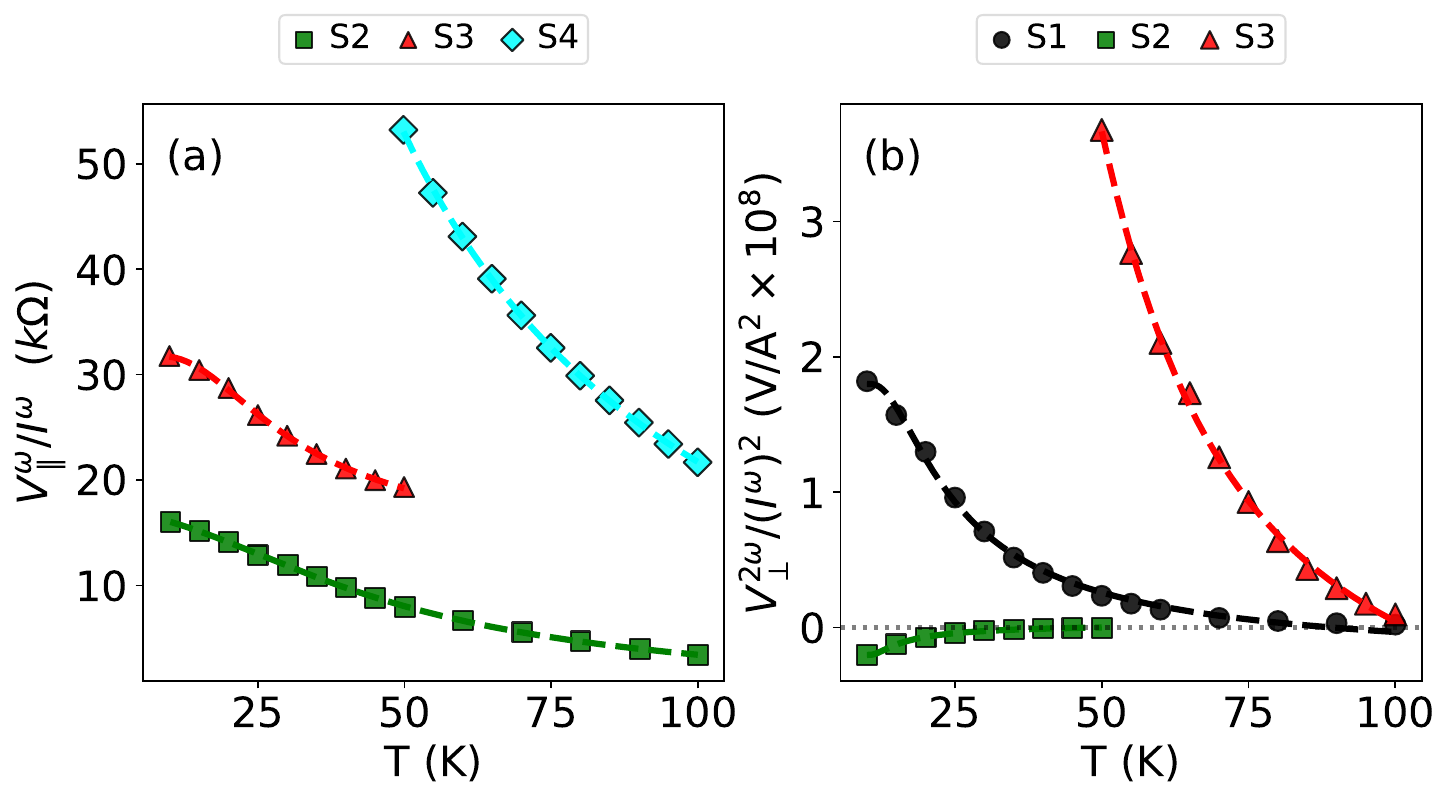}
\caption{
{Thermodynamics of lone-pair polarization.}
(a) First-harmonic $V^{\omega}/I^\omega$ compared with the prediction proportional to the second thermal moment $\langle P_x^2 \rangle_T$. 
(b) Second-harmonic $V^{2\omega}/(I^\omega)^2$ compared with the theoretical prediction proportional to the first thermal moment $\langle P_x \rangle_T$.}
\label{fig:modified-LD-model-three-samples}
\end{figure}

\textit{Temperature dependence.--}
Finally, we relate the temperature dependence of the harmonic signals to the thermodynamics of the surface lone-pair polarization.
The surface breaks the threefold equivalence of the local lone-pair directions, while preserving the residual reflection symmetry $2\leftrightarrow 3$. 
We can therefore write $w_{t,1}=1+2m_t/3$ and $w_{t,2}=w_{t,3}=1-m_t/3$, so that the net polarization becomes, $\mathbf{P}_t= p_0\sum_{a=1}^{3}w_{t,a}\hat{\mathbf{u}}_a=p_0m_t\hat{\mathbf{u}}_1$. 
Here $p_0$ is the microscopic dipole scale, and $m_t$ the
local imbalance between the surface-selected direction and the two remaining
equivalent directions. Keeping the three dominant local sectors
$m_t=+1,0,-1$, the single-triangle partition function is
\begin{equation}
    Z_t=1+e^{x(T)}+e^{-x(T)},\qquad
    x(T)=\frac{h_{\rm eff}(T)}{k_B T}.
\end{equation}
Here, $h_{\rm eff}(T)$ is the effective surface-induced polar field. It represents the boundary bias
created by the loss of coordination at the surface, dressed by correlations
between neighboring triangles and  thermal fluctuations (see \cite{SM} for details). 
The first two thermal moments of the local polarization are given by
\begin{equation}
    \langle P_x\rangle_T
    =
    p_0\frac{e^{x(T)}-e^{-x(T)}}{Z_t},
    \quad
    \langle P_x^2\rangle_T
    =
    p_0^2\frac{e^{x(T)}+e^{-x(T)}}{Z_t}.
\end{equation}
While $\langle P_x\rangle_T$ measures the net inversion-odd surface polarization, $\langle P_x^2\rangle_T$ quantifies the even polar strength, incorporating local polar fluctuations of either sign. 
The connection to transport directly follows from the quantum-geometric velocity framework.
Since the effective surface polar field is proportional to the lone-pair polarization
($\mathcal{E}_0 \propto P_x$), the first-harmonic correction $\delta\sigma_{xx}^{\rm geo} \propto \langle \mathcal{E}_0^2 \rangle_T$ is even in the polarization and scales with $\langle P_x^2\rangle_T$.
Conversely, the second-harmonic response $\chi_{xxx}^{\rm geo} \propto \langle \mathcal{E}_0 \rangle_T$ is inversion-odd  and tracks $\langle P_x\rangle_T$. 
As a result, quantum-geometric transport serves as a dual probe of both the fluctuating polar strength and the net inversion-odd surface order via distinct, measurable harmonic voltages (see Fig.~\ref{fig:modified-LD-model-three-samples}).

{\it Conclusion.--} 
The framework developed here opens several experimentally testable directions that extend beyond the interpretation of existing data. 
First, Eq.~\eqref{eq:vgeo_general_main} implies the existence of a higher-order, inversion-even nonlinear response that has not yet been reported, that is a third-harmonic longitudinal conductivity of the form
\begin{equation}
    \frac{V^{3\omega}}{(I^\omega)^3} \;\propto\; 
    \left(\frac{2\chi^2_{xxx}}{\sigma_D^5} - \frac{\zeta_{xxxx}}{\sigma_D^4}\right),
    \qquad \zeta_{xxxx} < 0,
\end{equation}
which leads to a characteristic gate-voltage dependence combining distinct power laws, $\propto \mu^{-6}$ and $\propto \mu^{-9/2}$. 
A second key direction concerns rectification. 
The nonlinear conductivity responsible for the second-harmonic response must also generate a rectified DC voltage under a driving AC or radiofrequency field $E_x(t)=E_{\rm RF}\cos\omega t$. 
Under open-circuit conditions, the induced voltage $V_x^{\rm DC}$ satisfies \cite{Suarez-Adv-Mat-Rectification}
\begin{equation}
    V_x^{\rm DC}\propto-\frac{\chi_{xxx}}{\sigma_D}\,E_{\rm RF}^2\propto\mathcal E_0\propto\langle P_x\rangle_T,
    \label{eq:VDC_rectification_conclusion}
\end{equation}
up to device-geometry and impedance-matching factors. 
This mechanism can be generalized to other materials featuring frustrated polar structures, surface-confined dipole textures, or geometrically constrained order parameters, 
Such systems represent ideal platforms for engineering QG rectifiers that can be tuned via strain, sample thickness, gate voltage, temperature, and surface design. 

\begin{acknowledgments}
N.N.B. acknowledges financial support from FAPES (Proc.~2023-V05FN). 
V.V. acknowledges financial support of PNRR MUR project PE0000023-NQSTI and PRIN 2022 (Prot.~20228YCYY7). 
This work is supported by: CNPq through the grants 442072/2023-6, 302557/2025-3, 444450/2024-6, 305227/2024-6, and 442781/2023-7, and FAPERJ through the grants E-26/210.100/2023 and E-26/210.781/2025. 
M.B.S.N. acknowledges INCT -- Advanced Quantum Materials, involving CNPq (Proc.~408766/2024-7), FAPESP (Proc.~2025/27091-3), and CAPES. M.S.-R., M.G., F.C. and L.E.H. acknowledge financial support by the Spanish MICIU/AEI/10.13039/501100011033 and by ERDF/EU (project numbers CEX2020-001038-M, PID2021-122511OB-I00, and PID2024-155708OB-I00). They are also supported by MICIU and by the European Union NextGenerationEU Plan (PRTR-C17.I1), and by the IKUR Strategy under the collaboration agreement between Donostia International Physics Center and CIC nanoGUNE on behalf of the Department of Education of the Basque Government.

\end{acknowledgments}


\bibliography{OddNLConductivity}


\clearpage
\onecolumngrid

\setcounter{equation}{0}
\setcounter{figure}{0}
\setcounter{table}{0}

\renewcommand{\theequation}{S\arabic{equation}}
\renewcommand{\thefigure}{S\arabic{figure}}
\renewcommand{\thetable}{S\arabic{table}}
\renewcommand{\thesection}{S\arabic{section}}

\begin{center}
{\Large \bf Supplementary Information}\\[0.5em]
{\large \bf Surface lone-pair polarization probed by quantum-geometric transport in tellurium}\\[1em]

Nathanael N. Batista,
Wendel S. Paz,
Manuel Suárez-Rodríguez,
Pierpaolo Fontana,
Victor Velasco,
Marcus V. O. Moutinho,
Chang Niu,
Peide D. Ye,
Marco Gobbi,
Fèlix Casanova,
Luis E. Hueso,
Caio Lewenkopf,
and Marcello B. Silva Neto
\end{center}

\vspace{1em}

\twocolumngrid

\section{Harmonics generated by $\mathbf j(\mathbf E)$}
\label{sec:harmonics_jE}

In this Section we derive the relation between the nonlinear conductivity defined under
electric-field bias and the harmonic voltages measured under current bias. We keep all
numerical factors and signs explicitly. Throughout this section, the amplitudes
$E^\omega$, $j^\omega$, $V^\omega$, $E^{2\omega}$, $j^{2\omega}$, and $V^{2\omega}$
denote real cosine amplitudes. Thus, for example,
\begin{equation}
    X(t)=X^\omega\cos\omega t+X^{2\omega}\cos 2\omega t+\cdots .
\end{equation}
Equivalently, if complex Fourier coefficients are used, the numerical factors below must
be converted accordingly.

We start from the local nonlinear current expansion
\begin{equation}
    j_a(t)=\sigma_{ab}E_b(t)+\chi_{abc}E_b(t)E_c(t)+O(E^3),
    \label{eq:SI_jE_tensor}
\end{equation}
where repeated Cartesian indices are summed. The conductivity tensor $\sigma_{ab}$ is the
linear conductivity, while $\chi_{abc}$ is the quadratic nonlinear conductivity. In the
main text the dominant nonlinear response is the longitudinal component $\chi_{xxx}$ in
the crystal frame. The crystallographic $z$-axis is the helical axis, and the special
in-plane polar axis selected by the surface lone-pair polarization is denoted by $x$.

For reference, let us first consider electric-field bias. If
\begin{equation}
    E_a(t)=E_a^\omega \cos\omega t,
\end{equation}
then the quadratic part of Eq.~\eqref{eq:SI_jE_tensor} is
\begin{eqnarray}
    j_a^{(2)}(t)|&=&
    \chi_{abc}E_b^\omega E_c^\omega \cos^2\omega t\nonumber\\
    &=&\frac{1}{2}\chi_{abc}E_b^\omega E_c^\omega
    +
    \frac{1}{2}\chi_{abc}E_b^\omega E_c^\omega\cos 2\omega t.
    \label{eq:SI_j2w_from_E}
\end{eqnarray}
Therefore, under voltage bias, the rectified and second-harmonic current amplitudes are
identical,
\begin{equation}
    j_a^{0}=
    \frac{1}{2}\chi_{abc}E_b^\omega E_c^\omega,
    \qquad
    j_a^{2\omega}=
    \frac{1}{2}\chi_{abc}E_b^\omega E_c^\omega.
    \label{eq:SI_j0_j2w}
\end{equation}
The experiments discussed in the main text, however, are performed under current bias.
One must therefore invert Eq.~\eqref{eq:SI_jE_tensor} and express the electric field as a
power series in the applied current.

\subsection{Current-bias inversion: from $\mathbf j(\mathbf E)$ to $\mathbf E(\mathbf j)$}

Let $\rho_{ab}$ be the inverse of the linear conductivity tensor,
\begin{equation}
    \rho_{ac}\sigma_{cb}=\delta_{ab}.
\end{equation}
We seek the perturbative inverse of Eq.~\eqref{eq:SI_jE_tensor} in the form
\begin{equation}
    E_a=\rho_{ab}j_b+\eta_{abc}j_bj_c+O(j^3).
    \label{eq:SI_Ej_ansatz}
\end{equation}
Substituting Eq.~\eqref{eq:SI_Ej_ansatz} into Eq.~\eqref{eq:SI_jE_tensor} gives
\begin{align}
    j_a
    &=
    \sigma_{ab}
    \left(
        \rho_{bc}j_c+\eta_{bcd}j_cj_d
    \right)
    +
    \chi_{abc}
    \left(
        \rho_{bd}j_d
    \right)
    \left(
        \rho_{ce}j_e
    \right)
    +O(j^3)
    \nonumber\\
    &=
    j_a
    +
    \left[
        \sigma_{ab}\eta_{bde}
        +
        \chi_{abc}\rho_{bd}\rho_{ce}
    \right]j_dj_e
    +O(j^3).
\end{align}
The quadratic term must vanish identically, which gives
\begin{equation}
    \eta_{ade}
    =
    -\rho_{ab}\chi_{bmn}\rho_{md}\rho_{ne}.
\end{equation}
Therefore, the exact tensorial inversion to second order in current is
\begin{equation}
    E_a
    =
    \rho_{ab}j_b
    -
    \rho_{ad}\chi_{def}\rho_{eb}\rho_{fc}j_bj_c
    +O(j^3).
    \label{eq:SI_Ej_tensor}
\end{equation}
This equation is the current-biased counterpart of the nonlinear conductivity expansion.
It shows that the measured second-harmonic electric field contains three factors of the
linear resistivity tensor, with an overall minus sign.

For a scalar isotropic conductor, Eq.~\eqref{eq:SI_Ej_tensor} reduces to
\begin{equation}
    E
    =
    \rho j
    -
    \chi \rho^3 j^2
    +O(j^3)
    =
    \frac{j}{\sigma}
    -
    \frac{\chi}{\sigma^3}j^2
    +O(j^3).
    \label{eq:SI_scalar_inversion}
\end{equation}
If the applied current is
\begin{equation}
    j(t)=j^\omega\cos\omega t,
\end{equation}
then
\begin{equation}
    E(t)
    =
    \rho j^\omega\cos\omega t
    -
    \chi\rho^3(j^\omega)^2\cos^2\omega t
    +O[(j^\omega)^3],
\end{equation}
or, equivalently,
\begin{align}
    \nonumber
    E(t)
    =&
    \rho j^\omega\cos\omega t
    -
    \frac{1}{2}\chi\rho^3(j^\omega)^2\\
    &-
    \frac{1}{2}\chi\rho^3(j^\omega)^2\cos 2\omega t
    +O[(j^\omega)^3].
\end{align}
Thus the rectified and second-harmonic electric-field amplitudes under current bias are
\begin{equation}
    E^{0}
    =
    -\frac{1}{2}\chi\rho^3(j^\omega)^2,
    \qquad
    E^{2\omega}
    =
    -\frac{1}{2}\chi\rho^3(j^\omega)^2.
    \label{eq:SI_scalar_E2w}
\end{equation}
The minus sign is important: a positive nonlinear conductivity produces a negative
second-harmonic electric field under current bias.

Converting from fields and current densities to voltages and currents,
\begin{equation}
    V^{2\omega}=L E^{2\omega},
    \qquad
    I^\omega=A j^\omega,
\end{equation}
{where $L$ is the effective length between voltage probes and $A$ is the cross-sectional area of the conducting channel,}
one obtains
\begin{equation}
    \frac{V^{2\omega}}{(I^\omega)^2}
    =
    -\frac{L}{2A^2}
    \chi\rho^3
    =
    -\frac{L}{2A^2}
    \frac{\chi}{\sigma^3}.
    \label{eq:SI_V2w_scalar_exact}
\end{equation}
The experimentally measured second-harmonic resistance therefore scales with
$-\chi\rho^3$, including the factor $1/2$ associated with
$\cos^2\omega t=(1+\cos2\omega t)/2$.

If the microscopic mechanism gives $\chi=C_\chi\tau$ and the Drude conductivity is
$\sigma=C_\sigma\tau$, with $C_\chi$ and $C_\sigma$ independent of the elastic lifetime
$\tau$, then Eq.~\eqref{eq:SI_V2w_scalar_exact} gives
\begin{equation}
    \frac{V^{2\omega}}{(I^\omega)^2}
    =
    -\frac{L}{2A^2}
    \frac{C_\chi}{C_\sigma^3}
    \tau^{-2}.
\end{equation}
Since $\rho=1/\sigma=(C_\sigma\tau)^{-1}$, this ration can also be written as 
\begin{equation}
    \frac{V^{2\omega}}{(I^\omega)^2}
    =
    -\frac{L}{2A^2}
    \frac{C_\chi C_\sigma^2}{C_\sigma^3}
    \rho^2
    =
    -\frac{L}{2A^2}
    \frac{C_\chi}{C_\sigma}
    \rho^2.
    \label{eq:SI_rho2_exact}
\end{equation}
Thus the $\rho^2$ scaling follows from the fact that the nonlinear conductivity contains
one power of $\tau$, whereas the current-biased voltage response contains three powers of
the resistivity.

\subsection{Angular dependence for dominant $\chi_{xxx}$}

We now specialize the tensor inversion to the experimental rotation geometry. The current
is injected in the $x-z$ plane. We define the longitudinal unit vector
\begin{equation}
    \hat{\mathbf e}_{\parallel}
    =
    \sin\theta\,\hat{\mathbf x}
    +
    \cos\theta\,\hat{\mathbf z}
    =
    (\sin\theta,0,\cos\theta),
    \label{eq:SI_eparallel}
\end{equation}
where $\theta$ is measured from the crystallographic $z-$axis. With this convention,
$\theta=0$ corresponds to current along the helical axis, while $\theta=90^\circ$
corresponds to current along the surface-polar axis $x$. The in-plane transverse direction
is chosen as the right-handed orthogonal vector
\begin{equation}
    \hat{\mathbf e}_{\perp}
    =
    \cos\theta\,\hat{\mathbf x}
    -
    \sin\theta\,\hat{\mathbf z}
    =
    (\cos\theta,0,-\sin\theta),
    \label{eq:SI_eperp}
\end{equation}
so that $\hat{\mathbf e}_{\parallel}\cdot\hat{\mathbf e}_{\perp}=0$, $|\hat{\mathbf e}_{\parallel}|=|\hat{\mathbf e}_{\perp}|=1$. The applied current density is therefore
\begin{equation}
    \mathbf j(t)
    =
    j^\omega\cos\omega t\,\hat{\mathbf e}_{\parallel},
\end{equation}
with Cartesian components
\begin{eqnarray}
    j_x(t)&=&j^\omega\sin\theta\cos\omega t,\nonumber\\
    j_z(t)&=&j^\omega\cos\theta\cos\omega t,\nonumber\\
    j_y(t)&=&0.
    \label{eq:SI_current_components}
\end{eqnarray}

We assume that the linear conductivity tensor is diagonal in the crystal frame,
\begin{equation}
    \sigma_{ab}
    =
    \begin{pmatrix}
        \sigma_{xx} & 0 & 0\\
        0 & \sigma_{yy} & 0\\
        0 & 0 & \sigma_{zz}
    \end{pmatrix},
    \qquad
    \rho_{ab}
    =
    \begin{pmatrix}
        \rho_{xx} & 0 & 0\\
        0 & \rho_{yy} & 0\\
        0 & 0 & \rho_{zz}
    \end{pmatrix},
    \label{eq:SI_diag_sigma}
\end{equation}
with $\rho_{ii}=1/\sigma_{ii}$. The first-order electric field generated by the current is
\begin{equation}
    \mathbf E^{(1)}(t)
    =
    \rho_{xx}j^\omega\sin\theta\cos\omega t\,\hat{\mathbf x}
    +
    \rho_{zz}j^\omega\cos\theta\cos\omega t\,\hat{\mathbf z}.
    \label{eq:SI_E1_aniso}
\end{equation}
Projecting Eq.~\eqref{eq:SI_E1_aniso} onto the current direction gives the usual angular
dependence of the linear longitudinal response,
\begin{align}
    E_{\parallel}^{\omega}
    &=
    \hat{\mathbf e}_{\parallel}\cdot\mathbf E^{\omega}
    \nonumber\\
    &=
    \rho_{xx}j^\omega\sin^2\theta
    +
    \rho_{zz}j^\omega\cos^2\theta.
\end{align}
Thus
\begin{equation}
    \frac{E_{\parallel}^{\omega}}{j^\omega}
    =
    \rho_{xx}\sin^2\theta
    +
    \rho_{zz}\cos^2\theta.
    \label{eq:SI_linear_angle}
\end{equation}

We now consider the second-order field produced by the dominant nonlinear tensor element
$\chi_{xxx}$. Keeping only this component in Eq.~\eqref{eq:SI_Ej_tensor}, we obtain
\begin{equation}
    E_a^{(2)}(t)
    =
    -\rho_{ax}\chi_{xxx}\rho_{xx}^2 j_x^2(t).
\end{equation}
Since the resistivity tensor is diagonal, only the $a=x$ component survives:
\begin{equation}
    \mathbf E^{(2)}(t)
    =
    E_x^{(2)}(t)\,\hat{\mathbf x},
    \qquad
    E_x^{(2)}(t)
    =
    -\chi_{xxx}\rho_{xx}^3 j_x^2(t).
    \label{eq:SI_E2_x_only}
\end{equation}
Using Eq.~\eqref{eq:SI_current_components},
\begin{eqnarray}
    j_x^2(t)&=&
    (j^\omega)^2\sin^2\theta\cos^2\omega t\nonumber\\
    &=&
    \frac{1}{2}(j^\omega)^2\sin^2\theta
    +
    \frac{1}{2}(j^\omega)^2\sin^2\theta\cos2\omega t.
\end{eqnarray}
Therefore the second-harmonic electric field generated by $\chi_{xxx}$ is
\begin{equation}
    \mathbf E^{2\omega}
    =
    -\frac{1}{2}
    \chi_{xxx}\rho_{xx}^3
    (j^\omega)^2
    \sin^2\theta\,
    \hat{\mathbf x}.
    \label{eq:SI_E2w_vector}
\end{equation}
This equation contains the whole angular structure before projection onto the measured
voltage direction. The factor $\sin^2\theta$ comes from the fact that the nonlinear
current-biased field is controlled by $j_x^2$, while the vector direction of the generated
field is fixed by the dominant tensor component $\chi_{xxx}$ and therefore points along
$\hat{\mathbf x}$.

The longitudinal second-harmonic field is the projection of
Eq.~\eqref{eq:SI_E2w_vector} onto $\hat{\mathbf e}_{\parallel}$:
\begin{align}
    E_{\parallel}^{2\omega}
    &=
    \hat{\mathbf e}_{\parallel}\cdot\mathbf E^{2\omega}
    \nonumber\\
    &=
    (\sin\theta\,\hat{\mathbf x}+\cos\theta\,\hat{\mathbf z})
    \cdot
    \left[
        -\frac{1}{2}
        \chi_{xxx}\rho_{xx}^3
        (j^\omega)^2
        \sin^2\theta\,
        \hat{\mathbf x}
    \right]
    \nonumber\\
    &=
    -\frac{1}{2}
    \chi_{xxx}\rho_{xx}^3
    (j^\omega)^2
    \sin^3\theta.
    \label{eq:SI_E2w_parallel}
\end{align}
Similarly, the transverse in-plane second-harmonic field is the projection onto
$\hat{\mathbf e}_{\perp}$:
\begin{align}
    E_{\perp}^{2\omega}
    &=
    \hat{\mathbf e}_{\perp}\cdot\mathbf E^{2\omega}
    \nonumber\\
    &=
    (\cos\theta\,\hat{\mathbf x}-\sin\theta\,\hat{\mathbf z})
    \cdot
    \left[
        -\frac{1}{2}
        \chi_{xxx}\rho_{xx}^3
        (j^\omega)^2
        \sin^2\theta\,
        \hat{\mathbf x}
    \right]
    \nonumber\\
    &=
    -\frac{1}{2}
    \chi_{xxx}\rho_{xx}^3
    (j^\omega)^2
    \sin^2\theta\cos\theta.
    \label{eq:SI_E2w_perp}
\end{align}
Equations~\eqref{eq:SI_E2w_parallel} and \eqref{eq:SI_E2w_perp} are the desired angular
laws. The longitudinal response contains three factors of $\sin\theta$: two from the
nonlinear generation through $j_x^2$, and one from projecting the generated
$\hat{\mathbf x}$-directed field back onto the current direction. The transverse response
contains the same $\sin^2\theta$ generation factor, but the projection onto the transverse
direction gives $\cos\theta$.

Finally, converting electric fields to voltages gives
\begin{equation}
    \frac{V_{\parallel}^{2\omega}}{(I^\omega)^2}
    =
    -\frac{L_{\parallel}}{2A^2}
    \chi_{xxx}\rho_{xx}^3
    \sin^3\theta,
    \label{eq:SI_V2w_parallel_exact}
\end{equation}
and
\begin{equation}
    \frac{V_{\perp}^{2\omega}}{(I^\omega)^2}
    =
    -\frac{L_{\perp}}{2A^2}
    \chi_{xxx}\rho_{xx}^3
    \sin^2\theta\cos\theta.
    \label{eq:SI_V2w_perp_exact}
\end{equation}
Here $A$ is the cross-sectional area through which the current flows, while
$L_{\parallel}$ and $L_{\perp}$ are the longitudinal and transverse voltage-probe
separations appropriate to the measured voltage drop. If the same voltage-probe distance
is used in both channels, then $L_{\parallel}=L_{\perp}$; otherwise the corresponding
geometrical factors must be kept explicitly.

In the nearly isotropic notation used in the main text, or when only the Drude
conductivity along the relevant polar direction is retained, one may set
$\rho_{xx}=\rho_D=1/\sigma_D$. Equations~\eqref{eq:SI_V2w_parallel_exact} and
\eqref{eq:SI_V2w_perp_exact} then become
\begin{eqnarray}
    \frac{V_{\parallel}^{2\omega}}{(I^\omega)^2}&=&
    -\frac{L_{\parallel}}{2A^2}
    \frac{\chi_{xxx}}{\sigma_D^3}
    \sin^3\theta,\nonumber\\
    \frac{V_{\perp}^{2\omega}}{(I^\omega)^2}\
    &=&
    -\frac{L_{\perp}}{2A^2}
    \frac{\chi_{xxx}}{\sigma_D^3}
    \sin^2\theta\cos\theta.
    \label{eq:SI_V2w_final}
\end{eqnarray}
The relative sign of the measured second-harmonic voltage therefore tracks the sign of
$-\chi_{xxx}$, together with the sign of the angular projection. In particular, at fixed
geometry and for $0<\theta<90^\circ$, a negative $\chi_{xxx}$ produces a positive
longitudinal second-harmonic voltage under the convention used above.

\subsection{Relation to the quantum-geometric correction}

In the quantum-geometric mechanism discussed in the main text, the surface lone-pair
polarization selects $\mathbf E_0=E_0\hat{\mathbf x}$ and generates a dominant nonlinear
conductivity $\chi_{xxx}^{\rm geo}$. Within the low-frequency relaxation-time
approximation,
\begin{equation}
    \chi_{xxx}^{\rm geo}
    =
    \frac{2e^4\tau E_0}{\hbar}
    \sum_n\int_{\mathbf k}
    \Gamma_{xxx}^{(n)}(\mathbf k)v_{x,n}(\mathbf k)
    \left[-\partial_\varepsilon f_n^{(0)}\right].
    \label{eq:SI_chi_geo}
\end{equation}
Substituting Eq.~\eqref{eq:SI_chi_geo} into
Eqs.~\eqref{eq:SI_V2w_parallel_exact} and
\eqref{eq:SI_V2w_perp_exact} gives
\begin{widetext}
\begin{eqnarray}
    \frac{V_{\parallel}^{2\omega}}{(I^\omega)^2}&=&
    -\frac{L_{\parallel}}{A^2}
    \frac{e^4\tau E_0}{\hbar}
    \rho_{xx}^3
    \left[
        \sum_n\int_{\mathbf k}
        \Gamma_{xxx}^{(n)}(\mathbf k)v_{x,n}(\mathbf k)
        \left(-\partial_\varepsilon f_n^{(0)}\right)
    \right]
    \sin^3\theta,
    \label{eq:SI_V2w_parallel_geo}\\
    \frac{V_{\perp}^{2\omega}}{(I^\omega)^2}&=&
    -\frac{L_{\perp}}{A^2}
    \frac{e^4\tau E_0}{\hbar}
    \rho_{xx}^3
    \left[
        \sum_n\int_{\mathbf k}
        \Gamma_{xxx}^{(n)}(\mathbf k)v_{x,n}(\mathbf k)
        \left(-\partial_\varepsilon f_n^{(0)}\right)
    \right]
    \sin^2\theta\cos\theta.
    \label{eq:SI_V2w_perp_geo}
\end{eqnarray}
\end{widetext}
These expressions show explicitly that the odd second-harmonic voltage is linear in the
surface polar scale $E_0$. Under inversion, $E_0\rightarrow -E_0$ and
$\chi_{xxx}^{\rm geo}\rightarrow -\chi_{xxx}^{\rm geo}$, so the measured
second-harmonic voltage changes sign. By contrast, the linear quantum-geometric
correction to the conductivity is even in $E_0$ and therefore does not change sign under
inversion.

If one writes the polar-axis conductivity as
\begin{equation}
    \sigma_{xx}
    =
    \sigma_D+\delta\sigma_{xx}^{\rm geo},
    \qquad
    |\delta\sigma_{xx}^{\rm geo}|\ll \sigma_D,
\end{equation}
then the corresponding resistivity is
\begin{eqnarray}
    \rho_{xx}&=&
    \frac{1}{\sigma_D+\delta\sigma_{xx}^{\rm geo}}
    =
    \rho_D
    -
    \frac{\delta\sigma_{xx}^{\rm geo}}{\sigma_D^2}
    +O[(\delta\sigma_{xx}^{\rm geo})^2],\nonumber\\
    \rho_D&=&\frac{1}{\sigma_D}.
    \label{eq:SI_rho_expand}
\end{eqnarray}
The first-harmonic longitudinal response therefore contains
\begin{equation}
    \frac{V_{\parallel}^{\omega}}{I^\omega}
    =
    \frac{L_{\parallel}}{A}
    \left[
        \rho_{zz}\cos^2\theta
        +
        \left(
            \rho_D
            -
            \frac{\delta\sigma_{xx}^{\rm geo}}{\sigma_D^2}
        \right)
        \sin^2\theta
    \right],
    \label{eq:SI_V1w_final}
\end{equation}
to first order in $\delta\sigma_{xx}^{\rm geo}$. Thus the geometric contribution to the
first-harmonic signal is proportional to
$-\delta\sigma_{xx}^{\rm geo}\sin^2\theta/\sigma_D^2$, while the second-harmonic
signals are controlled by
$-\chi_{xxx}\sin^3\theta/(2\sigma_D^3)$ and
$-\chi_{xxx}\sin^2\theta\cos\theta/(2\sigma_D^3)$ in the longitudinal and transverse
channels, respectively.


\section{Parity-odd semiclassical dynamics}
To clarify the role of the inversion-odd structure in the semiclassical description, we distinguish two distinct physical effects:
\begin{itemize}
    \item the external electric field, which controls the motion of the wavepacket in reciprocal space, and
    \item the intrinsic inversion-odd crystal potential, which does not act as a macroscopic force but instead modifies the internal structure of the Bloch wavepackets through inderband mixing effects.
\end{itemize}
Accordingly, only the external field enters the semiclassical equation of motion for the crystal momentum of the wavepacket, i.e.
\begin{equation}
    \hbar \dot{\mathbf k}_c = -e\,\mathbf E(t).
\end{equation}

\subsection{Microscopic origin of the polar scale}
We write the microscopic Hamiltonian as 
\begin{equation}
    \hat H=\hat H_0^{\rm even}+\hat H_0^{\rm odd}
    -e\,\mathbf E(t)\cdot \hat{\mathbf r},
    \label{eq:H_split_even_odd_ext}
\end{equation}
where $\hat H_0^{\rm even}$ contains the inversion-even part of the lattice
Hamiltonian, $\hat H_0^{\rm odd}$ contains the inversion-odd polar part of the crystal potential, and $\hat H_E(t)=-e\,\mathbf E(t)\cdot \hat{\mathbf r}$ is the coupling to the externally applied voltage field (here $e>0$ denotes the magnitude of the electron charge).

We denote by $\mathbf R_\ell$ the generic ionic position and let
\begin{equation}
    \boldsymbol\delta {\bf r} = \mathbf r - \mathbf R_\ell
\end{equation}
be the local coordinate measured from that site. The inversion-odd part of the crystal potential can be expanded locally as
\begin{equation}
    V_{\rm odd}(\mathbf R_\ell+\boldsymbol\delta {\bf r})=
    V_{\rm odd}(\mathbf R_\ell)+\left.\nabla V_{\rm odd}\right|_{\mathbf R_\ell}
    \cdot \boldsymbol\delta {\bf r}+\cdots .
    \label{eq:Vodd_local_expansion}
\end{equation}

In a centrosymmetric environment the linear term would be absent at the ionic
site. In a polar, inversion-odd environment, the local slope does not need to vanish, $\left.\nabla V_{\rm odd}\right|_{\mathbf R_\ell}\neq 0$. It is therefore useful to define a local polar dipolar scale $\boldsymbol{\mathcal E}_0$ as
\begin{equation}
    \left.
    \nabla V_{\rm odd}
    \right|_{\mathbf R_\ell}
    \cdot \boldsymbol\delta {\bf r}
    \equiv
    -e\,\boldsymbol{\mathcal E}_0\cdot\boldsymbol\delta {\bf r} .
    \label{eq:E0_definition}
\end{equation}
Equivalently,
\begin{equation}
    \boldsymbol{\mathcal E}_0=
    -\frac{1}{e}\left.\nabla V_{\rm odd}\right|_{\mathbf R_\ell}.
    \label{eq:E0_gradient_definition}
\end{equation}

Therefore, the leading dipolar contribution defines an intrinsic polar scale, which does not act as a macroscopic electric field, but as a compact parametrization of inversion-odd interband matrix elements.

\subsection{Unified dipolar coupling}
Looking at Eqs. \eqref{eq:H_split_even_odd_ext} and \eqref{eq:E0_definition}, both the external electric field and the intrinsic polar structure enter the wavepacket dynamics through the dipolar operator acting on the internal coordinate:
\begin{equation}
    \hat W_{\rm dip}
    =
    -e\left[\mathbf E(t)+\boldsymbol{\mathcal E}_0\right]
    \cdot(\hat{\mathbf r}-\mathbf r_c).
    \label{eq:total_dipolar_mixing_operator}
\end{equation}
However, $\boldsymbol{\mathcal E}_0$ is not a macroscopic electrostatic field and thus does not contribute to the semiclassical force equation.

We therefore define the effective dipolar field 
\begin{equation}
    \mathbf E_{\rm eff}=\mathbf E(t)+\boldsymbol{\mathcal E}_0
\end{equation}
which controls all interband mixing effects.

\subsection{Wavepacket geometry}
It has been shown that the first-order correction to a Bloch state due to an electromagnetic perturbation produces a gauge-invariant displacement of the wavepacket center of mass \cite{Gao2014}. On general ground, this positional shift reads
\begin{equation}
    \mathbf a_n'=\langle u_n|i\nabla_{\mathbf k}|u_n'\rangle
    +\langle u_n'|i\nabla_{\mathbf k}|u_n \rangle,
    \label{eq:positional_shift_definition}
\end{equation}
where $\ket{u_n}$ is the unperturbed cell-periodic Bloch function and
$\ket{u_n'}$ is its first-order correction due to the effective dipolar field, given by
\begin{equation}
    \ket{u_n'}=\sum_{m\neq n}\ket{u_m}
    \frac{\bra{u_m}\hat W_{\rm dip}\ket{u_n}}{\varepsilon_n-\varepsilon_m}.
    \label{eq:first_order_bloch_correction}
\end{equation}
Using the interband connections
\begin{equation}
    \mathcal A^i_{mn}=\langle u_m|i\partial_{k_i}|u_n\rangle,\qquad m\neq n,
\end{equation}
we can write the matrix element of the effective dipolar operator as
\begin{equation}
    \bra{u_m}\hat W_{\rm dip}\ket{u_n}=
    -e\left[E_i(t)+\mathcal E_{0,i}\right]\mathcal A^i_{mn}.
    \label{eq:dipole_matrix_element}
\end{equation}

For an external electric field and in the absence of magnetic field, this shift has the specific form
\begin{equation}
    a'_{n,i}=e\,\mathcal G^{(n)}_{ij} E_j ,
    \label{eq:gao_shift_external_field}
\end{equation}
where $\mathcal G^{(n)}_{ij}$ is the cross-gap geometric tensor
\begin{equation}
    \mathcal G^{(n)}_{ij}=2\,{\rm Re}\sum_{m\neq n}
    \frac{(v_i)_{nm}(v_j)_{mn}}{(\varepsilon_n-\varepsilon_m)^3}.
    \label{eq:gao_G_tensor}
\end{equation}
Here, $(v_i)_{nm}$ are interband velocity matrix elements. In a two-band model this tensor is directly proportional to the quantum metric with an additional energy-denominator factor. Thus, the positional shift is the real-space manifestation of quantum-geometric interband mixing. 

In analogy with Eq. \eqref{eq:gao_shift_external_field}, the positional shift in the present case becomes
\begin{equation}
    a'_{n,i}=e\,\mathcal G^{(n)}_{ij}\left[E_j(t)+\mathcal E_{0,j}\right].
    \label{eq:polar_positional_shift}
\end{equation}
This expression shows that the internal polar scale $\boldsymbol{\mathcal E}_0$ enters the real-space position of the wavepacket through the same cross-gap geometric tensor that governs the response to an external electric field.

\subsection{Energy correction}
By means of second order perturbation theory one can derive a correction to the unperturbed band energy. For a single isolated band $n$, it reads
\begin{equation}
    \Delta \tilde\varepsilon_n^{(2)}
    =
    e^2\Lambda^{(n)}_{ij} E_{\rm eff,i} E_{\rm eff,j},
    \label{eq:energy_correction_EplusE0}
\end{equation}
with $\Lambda^{(n)}_{ij}$ the quantum-geometric susceptibility of the band
\begin{equation}
    \Lambda^{(n)}_{ij}(\mathbf k)=
    \sum_{m\neq n}\frac{{\rm Re}\left[\mathcal A^i_{nm}\mathcal A^j_{mn}\right]}
    {\varepsilon_n-\varepsilon_m}.
    \label{eq:Lambda_metric_susceptibility}
\end{equation}
This susceptibility is closely related to the quantum metric
\begin{equation}
    g^{(n)}_{ij}(\mathbf k)=
    {\rm Re}\sum_{m\neq n}\mathcal A^i_{nm}\mathcal A^j_{mn},
    \label{eq:quantum_metric_definition}
\end{equation}
with an additional interband energy denominator. In an effective
two-band description, $\Lambda^{(n)}_{ij}$ reduces to the quantum metric divided by the band gap \cite{Kaplan_2024_PRL}.

Expanding Eq.~\eqref{eq:energy_correction_EplusE0} in terms of the external electric field and the polarization scale gives
\begin{align}
    \Delta\tilde\varepsilon_n^{(2)}&=
    e^2\Lambda^{(n)}_{ij}E_i(t)E_j(t)\nonumber\\
    &\quad
    +2e^2\Lambda^{(n)}_{ij}E_i(t)\mathcal E_{0,j}
    \nonumber\\
    &\quad
    +e^2\Lambda^{(n)}_{ij}\mathcal E_{0,i}\mathcal E_{0,j}.
    \label{eq:energy_expanded}
\end{align}
The three contributions have distinct roles. 
The first is the usual second-order Stark correction to the wavepacket energy, the last renormalizes the equilibrium polar band structure, and the mixed term encodes the key effect of interest: a response linear in the external field enabled by the intrinsic inversion-odd nontrivial structure.

\subsection{Semiclassical dynamics}
The semiclassical equations of motion retain their standard form \cite{Gao2014}
\begin{align}
    \hbar\dot{\mathbf k}_c &= -e\,\mathbf E(t), \\
    \dot{\mathbf r}_c &= \frac{1}{\hbar}\nabla_{\mathbf k}\tilde\varepsilon_n
    -\dot{\mathbf k}_c\times\widetilde{\boldsymbol\Omega}_n,
\end{align}
where
\begin{equation}
    \widetilde{\boldsymbol\Omega}_n=
    \boldsymbol\Omega_n+\boldsymbol\Omega_n',
    \qquad
    \boldsymbol\Omega_n'=\nabla_{\mathbf k}\times \mathbf a_n',
    \label{eq:corrected_berry_curvature}
\end{equation}
with $\mathbf a_n'$ is given by Eq. \eqref{eq:polar_positional_shift}. The quantum-geometric contribution to the band velocity is 
\begin{equation}
    \delta v_{n,\alpha} = \frac{e^2}{\hbar}
    \partial_{k_\alpha}\Lambda^{(n)}_{ij}E_{\rm eff,i}E_{\rm eff,j},
\end{equation}
which can be expanded on the same lines of Eq.~\eqref{eq:energy_expanded} to get the separate contributions 
\begin{align}
    \delta v_{n,\alpha}^{\rm geom}&=
    \frac{e^2}{\hbar}
    \partial_{k_\alpha}
    \Lambda^{(n)}_{ij}
    E_i(t)E_j(t)
    \nonumber\\
    &\quad
    +\frac{2e^2}{\hbar}\partial_{k_\alpha}\Lambda^{(n)}_{ij}E_i(t)\mathcal E_{0,j}
    \nonumber\\
    &\quad
    +\frac{e^2}{\hbar}
    \partial_{k_\alpha}
    \Lambda^{(n)}_{ij}
    \mathcal E_{0,i}\mathcal E_{0,j}.
    \label{eq:geometric_velocity_expanded}
\end{align}

The physical interpretation of the present construction is the following: the external electric field $\mathbf E(t)$ governs momentum-space transport via $\dot{\mathbf k}_c$, while the intrinsic polar structure acts only through interband mixing, modifying the wavepacket geometry, energy, and velocity. 
It does not introduce an additional semiclassical force, but instead enters transport observables through quantum-geometric response functions.

\section{L\"owdin downfolding and Christoffel symbols in the valence sector}
\label{sec:lowdin_christoffel_SI}

In this Section we derive the effective quantum-geometric structure of the
upper valence sector of tellurene. The goal is to obtain the Christoffel
symbols that enter the Fermi-surface average
\begin{equation}
    \left\langle v_a\Gamma_{bcd}\right\rangle_{\rm FS}
    =
    \sum_n
    \int\frac{d^2k}{(2\pi)^2}
    \left(-\frac{\partial f_0}{\partial\varepsilon_n}\right)
    v^n_a(\mathbf k)\,
    \Gamma^n_{bcd}(\mathbf k),
    \label{eq:vGamma_FS_def}
\end{equation}
where the motion is restricted to the experimentally relevant $k_x$--$k_z$
plane, so that $k_y=0$. We shall focus on the two components relevant for the
present nonlinear response,
\begin{equation}
    \Gamma_{xxx},
    \qquad
    \Gamma_{zxz}.
\end{equation}
The first one controls the longitudinal geodesic velocity generated by an
electric drive along the polar $x$ direction. The second one controls the
corresponding $z$-directed geodesic response involving one $x$ and one $z$
field component.

\subsection{Bare $H_4$--$H_5$ block}

We begin from the upper valence block
\begin{equation}
    \mathcal H_{45}(\mathbf k)
    =
    \begin{pmatrix}
        \varepsilon_{45}(\mathbf k)-S_1k_z-\Delta_2
        &
        -2\Delta_1+u_{45}(\mathbf k)
        \\
        -2\Delta_1+u^*_{45}(\mathbf k)
        &
        \varepsilon_{45}(\mathbf k)+S_1k_z-\Delta_2
    \end{pmatrix},
    \label{eq:H45_SI_lowdin}
\end{equation}
with
\begin{align}
    \varepsilon_{45}(\mathbf k)
    &=
    (A-r)k_\perp^2+B k_z^2,
    \\
    u_{45}(\mathbf k)
    &=
    -t k_\perp^2+2u k_z^2-2iw_1k_z .
\end{align}
For the leading geometric mechanism we may set $t=w_1=0$, as these parameters
are negligible in the valence-band parameter set \cite{Nakao_Doi_Kamimura_II}. The resulting two-band
Hamiltonian can be written as
\begin{equation}
    \mathcal H_{45}
    =
    d_0^{(45)}(\mathbf k)\,\mathbb I
    +
    \mathbf d_{45}(\mathbf k)\cdot\boldsymbol\sigma ,
\end{equation}
with, to leading order,
\begin{equation}
    \mathbf d_{45}(\mathbf k)
    =
    \left(
    -2\Delta_1+2u k_z^2,\,
    0,\,
    -S_1k_z
    \right).
    \label{eq:d45_bare_SI}
\end{equation}
Thus the bare $H_4$--$H_5$ block depends only on $k_z$ in its pseudospin texture. 
Consequently, although the quantum metric is finite, it has no {dependence on $k_x$}. 
In particular,
\begin{equation}
    \partial_{k_x}\tilde g_{xx}^{(45)}
    =
    \partial_{k_x}\tilde g_{zz}^{(45)}
    =
    0,
\end{equation}
and therefore
\begin{equation}
    \Gamma_{xxx}^{(45)}=0,
    \qquad
    \Gamma_{zxz}^{(45)}=0 .
    \label{eq:bare_Gamma_vanish_SI}
\end{equation}
This shows that an isolated $H_4$--$H_5$ model cannot generate the desired metric-dipole response. The finite Christoffel symbols must instead arise from virtual coupling to the remote $H_6,H_{6'}$ sector \cite{iacovelli2026polar}.

\subsection{L\"owdin downfolding of the $H_6,H_{6'}$ sector}

The full valence Hamiltonian is written in block form as
\begin{equation}
    \mathcal H_{4\times4}(\mathbf k)
    =
    \begin{pmatrix}
        \mathcal H_{45}(\mathbf k)
        &
        \mathcal T(\mathbf k)
        \\
        \mathcal T^\dagger(\mathbf k)
        &
        \mathcal H_{66'}(\mathbf k)
    \end{pmatrix}.
    \label{eq:H4x4_lowdin_SI}
\end{equation}
Since the $H_6,H_{6'}$ block is separated from the upper valence sector by the
large energy scale $\Delta_2$, we treat it perturbatively and use the
L\"owdin expression
\begin{equation}
    \mathcal H_{45}^{\rm eff}
    =
    \mathcal H_{45}
    -
    \mathcal T
    \left(
    \mathcal H_{66'}-\varepsilon
    \right)^{-1}
    \mathcal T^\dagger .
    \label{eq:Lowdin_general_SI}
\end{equation}
To leading order in $1/\Delta_2$,
\begin{equation}
    \left(
    \mathcal H_{66'}-\varepsilon
    \right)^{-1}
    \simeq
    \frac{1}{\Delta_2}\,\mathbb I ,
\end{equation}
so that
\begin{equation}
    \mathcal H_{45}^{\rm eff}
    \simeq
    \mathcal H_{45}
    -
    \frac{1}{\Delta_2}
    \mathcal T\mathcal T^\dagger .
    \label{eq:Lowdin_leading_SI}
\end{equation}

Keeping the leading hybridization terms that generate a new pseudospin
component, the product $\mathcal T\mathcal T^\dagger$ contains an identity
part, which only renormalizes the scalar dispersion, and a traceless
off-diagonal part. The identity contribution does not affect the quantum
metric. The traceless part generates an effective momentum-dependent
pseudospin component in the low-energy $H_4$--$H_5$ subspace. Restricting to
the $k_x$--$k_z$ plane, 
we obtain
\begin{equation}
    \mathcal H_{45}^{\rm eff}(\mathbf k)
    =
    d_0^{\rm eff}(\mathbf k)\,\mathbb I
    +
    \mathbf d^{\rm eff}(\mathbf k)\cdot\boldsymbol\sigma ,
\end{equation}
with
\begin{align}
    d_x^{\rm eff}(\mathbf k)
    &=
    M(k_z)
    \equiv
    -2\Delta_1+2u k_z^2,
    \\
    d_y^{\rm eff}(\mathbf k)
    &=
    -\beta\,k_x^2 k_z,
    \label{eq:dy_downfolded_SI}
    \\
    d_z^{\rm eff}(\mathbf k)
    &=
    -S_1 k_z,
\end{align}
where
\begin{equation}
    \beta \equiv \frac{2Rv}{\Delta_2}
    \label{eq:beta_def_SI}
\end{equation}
for real $v$. 
The essential result is Eq.~\eqref{eq:dy_downfolded_SI}: The remote Weyl-sector bands generate a $k_x^2k_z$ pseudospin texture in the otherwise trivial $H_4$--$H_5$ sector. 
This term is even in $k_x$ and odd in $k_z$, and it is 
{responsible for the} finite gradient of the quantum metric in the low-energy valence manifold.

For compactness, we define
\begin{equation}
    D(\mathbf k)
    \equiv
    |\mathbf d^{\rm eff}(\mathbf k)|
    =
    \left[
    M^2
    +
    S_1^2 k_z^2
    +
    \beta^2 k_x^4 k_z^2
    \right]^{1/2},
    \label{eq:D_eff_SI}
\end{equation}
where, in the last expression, the weak $k_z$ dependence of $M(k_z)$ may be
retained or neglected depending on the desired level of approximation. The
simplest analytic formulas below are written for slowly varying $M$, i.e.
$M\simeq -2\Delta_1$.

\subsection{Energy-normalized metric}

For a two-level Hamiltonian the quantum metric is
\begin{equation}
    g_{ij}
    =
    \frac{1}{4D^4}
    \left[
    D^2\,
    \partial_{k_i}\mathbf d\cdot\partial_{k_j}\mathbf d
    -
    \left(\mathbf d\cdot\partial_{k_i}\mathbf d\right)
    \left(\mathbf d\cdot\partial_{k_j}\mathbf d\right)
    \right].
    \label{eq:metric_two_band_SI}
\end{equation}
The geometric velocity entering the nonlinear response is controlled not by $g_{ij}$ itself, but by the energy-normalized metric
\begin{equation}
    \tilde g_{ij}
    =
    \frac{g_{ij}}{2D}.
    \label{eq:energy_normalized_metric_SI}
\end{equation}
This normalization accounts for the interband energy denominator associated
with the virtual transition between the two effective valence bands.

Using Eq.~\eqref{eq:metric_two_band_SI} with the downfolded vector above, and
setting $M$ slowly varying, one finds
\begin{align}
    g_{xx}
    &=
    \frac{
    \beta^2 k_x^2 k_z^2
    \left(M^2+S_1^2k_z^2\right)
    }
    {
    \left(
    M^2+S_1^2k_z^2+\beta^2k_x^4k_z^2
    \right)^2
    },
    \label{eq:gxx_downfolded_SI}
    \\
    g_{zz}
    &=
    \frac{
    M^2
    \left(
    S_1^2+\beta^2k_x^4
    \right)
    }
    {
    4\left(
    M^2+S_1^2k_z^2+\beta^2k_x^4k_z^2
    \right)^2
    },
    \label{eq:gzz_downfolded_SI}
    \\
    g_{xz}
    &=
    \frac{
    M^2\beta^2 k_x^3 k_z
    }
    {
    2\left(
    M^2+S_1^2k_z^2+\beta^2k_x^4k_z^2
    \right)^2
    } .
    \label{eq:gxz_downfolded_SI}
\end{align}
These metric components are even under 
$\mathbf k\rightarrow-\mathbf k$, as required by time-reversal symmetry, but
they now carry a nontrivial $k_x$ dependence inherited from the downfolded
remote bands.

\subsection{Christoffel symbols $\Gamma_{xxx}$ and $\Gamma_{zxz}$}

The Christoffel symbols constructed from the energy-normalized quantum metric
are defined as
\begin{equation}
    \Gamma_{abc}
    =
    \frac12
    \left(
    \partial_{k_b}\tilde g_{ac}
    +
    \partial_{k_c}\tilde g_{ab}
    -
    \partial_{k_a}\tilde g_{bc}
    \right).
    \label{eq:Christoffel_SI}
\end{equation}
For the longitudinal symbol we obtain
\begin{equation}
    \Gamma_{xxx}
    =
    \frac12\,\partial_{k_x}\tilde g_{xx}.
    \label{eq:Gamma_xxx_def_SI}
\end{equation}
Substituting Eq.~\eqref{eq:gxx_downfolded_SI} into Eq.~\eqref{eq:energy_normalized_metric_SI} results that
\begin{equation}
    \Gamma_{xxx}(\mathbf k)
    =
    \frac{
    \beta^2 k_x k_z^2
    \left(M^2+S_1^2k_z^2\right)
    \left[
    M^2+S_1^2k_z^2
    -
    4\beta^2 k_x^4 k_z^2
    \right]
    }
    {
    2
    \left(
    M^2+S_1^2k_z^2+\beta^2k_x^4k_z^2
    \right)^{7/2}
    } .
    \label{eq:Gamma_xxx_final_SI}
\end{equation}
This component is odd in $k_x$ and even in $k_z$. 
Consequently, the product $v_x\Gamma_{xxx}$ is even on the Fermi surface, because $v_x$ is also odd in $k_x$ in the parabolic valence-band approximation.

The second component required in the present geometry is
\begin{equation}
    \Gamma_{zxz}
    =
    \frac12
    \left(
    \partial_{k_x}\tilde g_{zz}
    +
    \partial_{k_z}\tilde g_{zx}
    -
    \partial_{k_z}\tilde g_{xz}
    \right).
\end{equation}
Since $\tilde g_{xz}=\tilde g_{zx}$, the last two terms identically cancel 
\begin{equation}
    \Gamma_{zxz}
    =
    \frac12\,\partial_{k_x}\tilde g_{zz}.
    \label{eq:Gamma_zxz_def_SI}
\end{equation}
Using Eq.~\eqref{eq:gzz_downfolded_SI}, one obtains
\begin{equation}
    \Gamma_{zxz}(\mathbf k)
    =
    \frac{
    M^2\beta^2 k_x^3
    \left[
    2M^2
    -
    3S_1^2k_z^2
    -
    3\beta^2k_x^4k_z^2
    \right]
    }
    {
    8
    \left(
    M^2+S_1^2k_z^2+\beta^2k_x^4k_z^2
    \right)^{7/2}
    } .
    \label{eq:Gamma_zxz_final_SI}
\end{equation}
This component is also odd in $k_x$. 
Thus, $v_z\Gamma_{zxz}$ is odd in $k_z$ and does not contribute to a strictly inversion-symmetric Fermi-surface average unless the remaining factors in the nonlinear transport kernel supply
the required odd-in-$k_z$ structure. 
In the experimental tensor projection, however, $\Gamma_{zxz}$ remains the natural Christoffel component associated with a $z$-directed geodesic response driven by mixed $x$-$z$ acceleration.

\subsection{Scaling of the Fermi-surface average}
\label{subsec:vGamma_scaling_SI}

We now extract the chemical-potential scaling of the Fermi-surface average that enters the geometric nonlinear response. 
{The starting point is the full electric magnetochiral anisotropy tensor written as}
\begin{equation}
    G_{ijk\ell}
    \sim
    \int\frac{d^2k}{(2\pi)^2}\,
    v_i(\mathbf k)\,
     \Gamma_{ak\ell}(\mathbf k)\,
    \partial_a v_j(\mathbf k)\,
    \left(
    -\frac{\partial f_0}{\partial\varepsilon}
    \right),
    \label{eq:G_scaling_schematic_SI}
\end{equation}
where $\partial_a v_j$ is constant within the parabolic-band approximation. 
The scaling is controlled by the product $v\,\Gamma$ evaluated at the Fermi surface. 
Therefore, we treat as the relevant object {the} two-dimensional momentum integral {defined in Eq.~\eqref{eq:vGamma_FS_def}}, 
up to tensorial prefactors, velocity derivatives, Levi-Civita symbols, and the polarization field $\mathcal E_{0,i}$, which do not affect the scaling argument.

We set the zero of energy at the upper valence-band maximum and define the positive hole energy
\begin{equation}
    \varepsilon \equiv |\mu|=-\mu>0 .
\end{equation}
For the upper valence band, we use the parabolic approximation
\begin{equation}
    \varepsilon_{\mathbf k}
    =
    -\frac{\hbar^2 k_x^2}{2m_x}
    -
    \frac{\hbar^2 k_z^2}{2m_z}.
    \label{eq:parabolic_valence_scaling_SI}
\end{equation}
Equivalently, after rescaling momenta to absorb the mass anisotropy, the
Fermi surface is a circle defined by
\begin{equation}
    \bar k^2
    =
    k_F^2,
    \qquad
    k_F^2
    \propto
    \varepsilon .
    \label{eq:kF_mu_relation_SI}
\end{equation}
Therefore $k_F\sim \varepsilon^{1/2}$. At low temperature,
\begin{equation}
    -\frac{\partial f_0}{\partial\varepsilon}
    \longrightarrow
    \delta(\varepsilon_{\mathbf k}-\mu).
\end{equation}
Using polar coordinates in the rescaled two-dimensional momentum plane,
\begin{equation}
    d^2k
    =
    k\,dk\,d\varphi ,
\end{equation}
the Fermi-surface integral becomes
\begin{align}
    \left\langle v\,\Gamma\right\rangle_{\rm FS}
    &\sim
    \int_0^{2\pi}d\varphi
    \int_0^\infty k\,dk\,
    v_i(k,\varphi)\,
    \Gamma_{abc}(k,\varphi)\,
    \delta\!\left(Ak^2-\varepsilon\right),
    \label{eq:vGamma_radial_integral_SI}
\end{align}
where $A$ is a positive constant determined by the effective masses. The
radial integral is then evaluated using
\begin{equation}
    \delta(Ak^2-\varepsilon)
    =
    \frac{1}{2Ak_F}\,
    \delta(k-k_F),
    \label{eq:delta_radial_identity_SI}
\end{equation}
which gives
\begin{equation}
    \int_0^\infty k\,dk\,
    F(k,\varphi)\,
    \delta(Ak^2-\varepsilon)
    =
    \frac{1}{2A}
    F(k_F,\varphi).
    \label{eq:radial_delta_result_SI}
\end{equation}
{Here, $F(k,\varphi)$ denotes the combination of momentum-dependent factors in the integrand, namely $F(k,\varphi)=v_i(k,\varphi)\Gamma_{abc}(k,\varphi)$.} 
Thus, the explicit factor of $k$ from the two-dimensional measure is canceled by the Jacobian of the delta function. 
The scaling is 
obtained by evaluating the remaining integrand directly on the Fermi surface.

The velocity scales as
\begin{equation}
    v_i(k_F)
    =
    \frac{1}{\hbar}
    \partial_{k_i}\varepsilon_{\mathbf k}
    \sim
    k_F
    \sim
    \varepsilon^{1/2}.
    \label{eq:v_scaling_SI}
\end{equation}
The nontrivial part is represented by the Christoffel symbol. From the downfolded
Hamiltonian,
\begin{equation}
    \mathbf d^{\rm eff}(\mathbf k)
    =
    \left(
    M,\,
    -\beta k_x^2k_z,\,
    -S_1k_z
    \right),
\end{equation}
[with Eqs.~\eqref{eq:Gamma_xxx_final_SI} and \eqref{eq:Gamma_zxz_final_SI}] one sees that the Christoffel symbols have the common denominator
\begin{equation}
    D^7
    =
    \left(
    M^2
    +
    S_1^2k_z^2
    +
    \beta^2k_x^4k_z^2
    \right)^{7/2}.
    \label{eq:D7_scaling_SI}
\end{equation}
For the scaling regime relevant to the asymptotic hole-doped valence band, we consider momenta large enough that the constant gap scale is no longer the dominant term,
\begin{equation}
    S_1^2k_z^2
    \gg
    M^2 ,
    \label{eq:large_k_gap_negligible_SI}
\end{equation}
while the higher-order $\beta^2k_x^4k_z^2$ term only modifies the angular
dependence and numerical prefactor. In this regime $D^7 \sim k_F^7$.

The numerator of the relevant Christoffel symbols contains five powers of momentum in the same asymptotic regime. 
More explicitly, for the longitudinal component {in Eq. \eqref{eq:Gamma_xxx_final_SI}},
dropping the gap scale $M$ for power counting, the leading part behaves as
\begin{equation}
    \Gamma_{xxx}(k_F,\varphi)
    \sim
    \frac{k_F^5}{k_F^7}
    =
    k_F^{-2},
    \label{eq:Gamma_xxx_power_count_SI}
\end{equation}
up to an angular function of $\varphi$. 
The same power counting applies to the transverse Christoffel component, given by Eq. \eqref{eq:Gamma_zxz_final_SI}, namely
\begin{equation}
    \Gamma_{zxz}(k_F,\varphi)
    \sim
    \frac{k_F^5}{k_F^7}
    =
    k_F^{-2},
    \label{eq:Gamma_zxz_power_count_SI}
\end{equation}
again up to angular factors and possible parity cancellations depending on the tensor component being integrated.

Combining Eqs.~\eqref{eq:radial_delta_result_SI},
\eqref{eq:v_scaling_SI}, and \eqref{eq:Gamma_xxx_power_count_SI}, the
Fermi-surface average scales as
\begin{equation}
    \left\langle v\,\Gamma\right\rangle_{\rm FS}\sim v(k_F)\,
    \Gamma(k_F)\sim
    k_F^{-1},
    \label{eq:vGamma_kF_scaling_SI}
\end{equation}
{and,} since $k_F\sim\varepsilon^{1/2}$, this gives
\begin{equation}
    \left\langle v\,\Gamma\right\rangle_{\rm FS}
    \propto\varepsilon^{-1/2}=|\mu|^{-1/2}.
    \label{eq:vGamma_mu_scaling_SI}
\end{equation}

\subsection{Symbols relevant for the transverse (Hall-like) nonlinear response}
\label{subsec:transverse_hall_symbols}

The transverse second-harmonic voltage measured in the $x-z$ plane can arise from two distinct mechanisms. The first is a genuine transverse nonlinear conductivity channel, e. g. a microscopic contribution to $\chi_{zxx}$. The second is purely geometric at the level of the measured voltage: even if the intrinsic nonlinear conductivity is dominated by $\chi_{xxx}$, tensor inversion under current bias and projection onto the transverse voltage probes generate a transverse signal with the angular form
$\sin^2\theta\cos\theta$.

The genuine transverse channel is controlled by Christoffel symbols that generate a $z$-directed geodesic velocity under in-plane driving. 
In the present $k_y=0$ geometry, the relevant components are $\Gamma_{zxx}$ and $\Gamma_{zxz}$, together with symmetry-related components. 
{Recalling Eq.~\eqref{eq:Christoffel_SI}, we obtain}
\begin{align}
    \Gamma_{zxx}
    &=
    \partial_{k_x}\tilde g_{zx}
    -
    \frac12\partial_{k_z}\tilde g_{xx},
    \label{eq:Gamma_zxx_def_SI}
    \\
    \Gamma_{zxz}
    &=
    \frac12
    \left(
    \partial_{k_x}\tilde g_{zz}
    +
    \partial_{k_z}\tilde g_{zx}
    -
    \partial_{k_z}\tilde g_{xz}
    \right)
    =
    \frac12\partial_{k_x}\tilde g_{zz},
    \label{eq:Gamma_zxz_transverse_def_SI}
\end{align}
where we have used that $\tilde g_{xz}=\tilde g_{zx}$.


Substituting the quantum-metric components given by Eqs.~\eqref{eq:gxx_downfolded_SI}, \eqref{eq:gzz_downfolded_SI}, and \eqref{eq:gxz_downfolded_SI} into the Löwdin-downfolded effective Hamiltonian in the $k_x$-$k_z$ plane yields the compact result 
\begin{equation}
    \Gamma_{zxz}(\mathbf k)=
    \frac{M^2\beta^2 k_x^3\left[2M^2-3S_1^2k_z^2-3\beta^2k_x^4k_z^2\right]}
    {8\left(M^2+S_1^2k_z^2+\beta^2k_x^4k_z^2\right)^{7/2}} .
    \label{eq:Gamma_zxz_transverse_final_SI}
\end{equation}
This symbol is odd in $k_x$, hence it contributes to a Fermi-surface average only when multiplied by another odd-in-$k_x$ factor, or when the experimental projection selects the corresponding parity-allowed tensor component.

The other transverse symbol is
\begin{widetext}
\begin{equation}
    \Gamma_{zxx}(\mathbf k)
    =
    \partial_{k_x}
    \left[
    \frac{
    M^2\beta^2 k_x^3k_z
    }
    {
    4
    \left(
    M^2+S_1^2k_z^2+\beta^2k_x^4k_z^2
    \right)^{5/2}
    }
    \right]
    -
    \frac12
    \partial_{k_z}
    \left[
    \frac{
    \beta^2 k_x^2 k_z^2
    \left(M^2+S_1^2k_z^2\right)
    }
    {
    2
    \left(
    M^2+S_1^2k_z^2+\beta^2k_x^4k_z^2
    \right)^{5/2}
    }
    \right] .
    \label{eq:Gamma_zxx_transverse_final_SI}
\end{equation}
\end{widetext}
We leave Eq.~\eqref{eq:Gamma_zxx_transverse_final_SI} {without computing the derivatives} because it makes the two geometric origins transparent: the first term is the $x$-gradient of the mixed metric $\tilde g_{xz}$, while the second term is the $z$-gradient of the longitudinal metric $\tilde g_{xx}$. Both arise only after the L\"owdin downfolding, which generates the $d_y^{\rm eff}\propto k_x^2k_z$ texture.

It is important to emphasize that in the minimal cylindrically symmetric approximation, the intrinsic transverse Fermi-surface average associated with $\Gamma_{zxx}$ or $\Gamma_{zxz}$ can be
strongly suppressed, or even vanish, by parity after integration over the closed Fermi contour. This does not imply that the measured transverse second-harmonic voltage must vanish. When the dominant microscopic nonlinear conductivity is $\chi_{xxx}$, the current-bias inversion and the projection of the resulting nonlinear electric field onto the transverse probe direction produce
\begin{equation}
    V_\perp^{2\omega}
    \propto
    \sin^2\theta\cos\theta ,
\end{equation}
even in the absence of a large intrinsic $\chi_{zxx}$. We therefore interpret
the transverse voltage primarily as a projection diagnostic of the dominant
$\chi_{xxx}$ channel, while $\Gamma_{zxx}$ and $\Gamma_{zxz}$ describe
additional intrinsic Hall-like channels that may become visible when
crystalline anisotropies beyond the simplest $k_\perp^2$ model are retained.

\subsection{Summary: interband origin of quantum geometry in $H_{45}$}
\label{subsec:summary_interband_geometry_H45}

The bare $H_4$--$H_5$ valence block is not sufficient to generate the
quantum-geometric nonlinear response required here. In the minimal
approximation, its pseudospin texture depends only on $k_z$, and therefore its energy-normalized quantum metric does not have a leading $k_x$-gradient. As a result, the relevant Christoffel symbols are $\Gamma_{xxx}^{(45)}=0,$ and  $\Gamma_{zxz}^{(45)}=0$ \cite{iacovelli2026polar}. Therefore, the isolated $H_{45}$ sector cannot produce the desired Fermi-surface average $\langle v\Gamma\rangle_{\rm FS}$.

The essential physics enters through virtual transitions to the remote $H_6,H_{6'}$ bands, since L\"owdin downfolding generates an effective pseudospin texture and the low-energy valence band inherits a quantum geometry from these remote Weyl-sector bands. The term $d_y^{\rm eff}\propto k_x^2k_z$ in Eq.  \eqref{eq:dy_downfolded_SI} is the minimal ingredient that produces finite metric gradients and, therefore, finite Christoffel symbols.

The two most relevant Christoffel symbols for the present nonlinear response are the ones written in Eqs. \eqref{eq:Gamma_xxx_final_SI}) and \eqref{eq:Gamma_zxz_final_SI}. The first one controls the dominant longitudinal nonlinear channel $\chi_{xxx}^{\rm geo}$, whereas the other component enters the possible intrinsic transverse, Hall-like channel. In the experimental geometry,
however, a transverse second-harmonic voltage can also be generated by tensor inversion and projection of the dominant $\chi_{xxx}$ response, giving the robust angular dependence
\begin{equation}
    V_\parallel^{2\omega}\propto \sin^3\theta,
    \qquad
    V_\perp^{2\omega}\propto \sin^2\theta\cos\theta .
\end{equation}

Finally, within the parabolic approximation for the valence band, the
Fermi-surface momentum scales as $k_F\sim |\mu|^{1/2}$, while the downfolded
Christoffel symbols carry inverse powers of the same momentum scale in the
large-hole-density regime. Combining this with the band velocity
$v\sim k_F$, one obtains
\begin{equation}
    \left\langle v\Gamma\right\rangle_{\rm FS}
    \sim
    |\mu|^{-1/2}.
\end{equation}
This scaling is the 
analytic signature of the present valence-band mechanism: the nonlinear response is generated not by a Weyl node at the Fermi surface, but by the quantum geometry inherited through interband L\"owdin dressing of the otherwise trivial $H_{45}$ valence manifold.

\section{Derivation of the first and second thermal moments from the coarse-grained polar lattice model}
In this Section we derive the expressions used in the main text for the first and second thermal moments of the surface-induced lone-pair polarization.

\subsection{Reduction of the coarse-grained energy to a single imbalance variable}

We begin from the coarse-grained lattice energy
\begin{align}
    \nonumber
    E =& \frac{\kappa}{2}\sum_{t,a}(w_{t,a}-1)^2
    +
    \frac{J}{2}\sum_{(t\rightarrow n,a)}(w_{t,a}-w_{n,a})^2
    \\
    &-
    \sum_t h(j_t)\, w_{t,1},
    \label{eq:Ecoarse_SI}
\end{align}
where $t$ labels triangles, $a=1,2,3$ labels the three local lone-pair directions, $J$ is the inter-triangle matching energy, $\kappa$ restores bulk equivalence, and $h(j_t)$ is the boundary-induced bias favoring the $a=1$ direction.

{As emphasized in the main text, close to} the surfaces the residual symmetry between directions $2$ and $3$ implies
\begin{equation}
w_{t,2}=w_{t,3}.
\end{equation}
It is then convenient to parametrize the three weights by a single scalar imbalance variable $m_t$:
\begin{equation}
w_{t,1}=1+\frac{2m_t}{3},\qquad
w_{t,2}=w_{t,3}=1-\frac{m_t}{3}.
\label{eq:wparam}
\end{equation}
This choice preserves the average weight
\begin{equation}
\frac{w_{t,1}+w_{t,2}+w_{t,3}}{3}=1
\end{equation}
while allowing an imbalance between direction $1$ and the pair $(2,3)$. Using the identity
\begin{equation}
\hat u_1+\hat u_2+\hat u_3=0,
\end{equation}
the triangle dipole moment becomes
\begin{equation}
\mathbf P_t
=
p_0\sum_{a=1}^3 w_{t,a}\hat u_a
=
p_0(w_{t,1}-w_{t,2})\hat u_1
=
p_0 m_t \hat u_1.
\label{eq:Pt_mt}
\end{equation}
Thus the polarization is completely controlled by the scalar variable $m_t$.

\subsection{Effective local three-sector description}

To obtain a minimal thermodynamic description \cite{cardy}, we retain the three dominant local sectors
\begin{equation}
m_t=+1,\qquad m_t=0,\qquad m_t=-1,
\label{eq:three_sector}
\end{equation}
corresponding respectively to:
\begin{itemize}
\item a locally positive polarization along $\hat x$,
\item a bulk-like locally compensated state,
\item a locally negative polarization along $\hat x$.
\end{itemize}

In terms of Eq.~\eqref{eq:Pt_mt}, these three sectors carry local polarization
\begin{equation}
P_{x,t}=P_0 s_t,
\qquad
s_t\in\{+1,0,-1\},
\end{equation}

The role of the interaction terms in Eq.~\eqref{eq:Ecoarse_SI} is twofold:
\begin{enumerate}
\item they penalize local deviations from bulk equivalence {($\propto \kappa$)},
\item they renormalize the response to the boundary field through inter-triangle correlations {($\propto J$)}.
\end{enumerate}
At the level of a single-site mean-field treatment, these effects can be absorbed into an effective local field $h_{\rm eff}(T)$, so that the effective local energy reads
\begin{equation}
E_{\rm loc}(s)=-h_{\rm eff}(T)\,s,
\qquad
s\in\{+1,0,-1\}.
\label{eq:Eloc}
\end{equation}

We define the corresponding dimensionless field
\begin{equation}
x(T)\equiv \beta h_{\rm eff}(T),
\qquad
\beta=\frac{1}{k_B T}.
\label{eq:xdef}
\end{equation}
More generally, to incorporate Weiss-like interaction effects, one may regard $x(T)$ as the dimensionless ratio
\begin{equation}
x(T)\propto \frac{h(T)}{T+\theta},
\label{eq: x(T)}
\end{equation}
where $\theta$ encodes the correlation-induced renormalization of the local response.

{Finally, the local partition function for the considered three-state statistical model}  {is written as} 
\begin{equation}
    Z
    =
    \sum_{s=0,\pm1} e^{-\beta E_{\rm loc}(s)}
    =
    1+e^{x(T)}+e^{-x(T)}.
    \label{eq:Zlocal}
\end{equation}

\subsection{First moment: thermal average polarization}

The thermal average of the local polarization is
\begin{equation}
\langle P_x\rangle_T
=
\frac{1}{Z}\sum_{s=0,\pm1} (P_0 s)\,e^{-\beta E_{\rm loc}(s)}.
\end{equation}
Substituting Eq.~\eqref{eq:Eloc} gives
\begin{equation}
    \langle P_x\rangle_T
    =
    P_0\,
    \frac{e^{x(T)}-e^{-x(T)}}{1+e^{x(T)}+e^{-x(T)}}.
    \label{eq:Px_exact_SI}
\end{equation}

This is the odd moment of the polarization, which changes sign under inversion of the polar axis, $x(T)\to -x(T)$.

\subsection{Second moment: even polar sector}

The second moment is
\begin{equation}
    \langle P_x^2\rangle_T
    =
    \frac{1}{Z}\sum_{s=0,\pm1} (P_0^2 s^2)\,e^{-\beta E_{\rm loc}(s)}.
\end{equation}
Since $s^2=1$ for $s=\pm1$ and $s^2=0$ for $s=0$, this becomes

\begin{equation}
    \langle P_x^2\rangle_T
    =
    P_0^2\,
    \frac{e^{x(T)}+e^{-x(T)}}{1+e^{x(T)}+e^{-x(T)}}.
    \label{eq:P2_exact_SI}
\end{equation}

This is the even sector of the polarization, as it is invariant under inversion, $x(T)\to -x(T)$.

\subsection{High-temperature (weak-bias) limit and Curie--Weiss form}

For weak bias, i.e. $x(T)\ll1$, we {expand the exponential weights in Eq.~\eqref{eq:Px_exact_SI} to obtain}
\begin{equation}
    \langle P_x\rangle_T
    =
    P_0\frac{2x(T)+O(x^3)}{3+O(x^2)}
    \simeq
    \frac{2P_0}{3}x(T).
\end{equation}
Thus, using Eq. (\ref{eq: x(T)}), we recover the Curie--Weiss form
\begin{equation}
    \langle P_x\rangle_T
    \propto
    \frac{h(T)}{T+\theta}.
    \label{eq:CWrecover}
\end{equation}

This is the origin of the fitting form used in the main text,
\begin{equation}
    h(T)=h_0\left(1-e^{-T/\tau}\right),
    \qquad
    \langle P_x\rangle_T \propto \frac{1-e^{-T/\tau}}{T+\theta}.
\end{equation}

\subsection{High-temperature limit of the second moment}

In the same weak-bias limit,
\begin{equation}
\langle P_x^2\rangle_T
=
P_0^2\frac{2+O(x^2)}{3+O(x^2)}
\simeq
\frac{2}{3}P_0^2 + O(x^2).
\end{equation}
Therefore the raw second moment approaches a finite constant as $T\to\infty$.

This is not a contradiction: the first-harmonic transport does not necessarily couple to the absolute second moment, but rather to its temperature-dependent excess over the isotropic background. Accordingly, the quantity relevant for fitting the experimental anisotropy is
\begin{equation}
\Delta\langle P_x^2\rangle_T
\equiv
\langle P_x^2\rangle_T-\langle P_x^2\rangle_{x=0},
\end{equation}
or, equivalently, the same expression supplemented by an additive offset parameter in the fitting function.

\subsection{Transport interpretation}

The two moments derived above map directly onto the two experimentally observed channels:
\begin{equation}
    \frac{V^{2\omega}}{(I^\omega)^2}
    \propto
    \langle P_x\rangle_T,
\end{equation}
\begin{equation}
    \frac{V^\omega}{I^\omega}
    \propto
    \Delta\langle P_x^2\rangle_T
    \quad
    (\text{or } \langle P_x^2\rangle_T \text{ plus a constant offset}).
\end{equation}
Thus the odd nonlinear response probes the first moment of the boundary-induced polarization, whereas the even linear anisotropy probes its second moment.

\section{Computational Methods}

\subsection{Density Functional Theory Calculations}
All slab calculations were performed within density functional theory 
(DFT) using norm-conserving pseudopotentials and pseudo-atomic 
localized basis functions (LCPAO), as implemented in the OpenMX 
package~\cite{Ozaki2003, OzakiKino2004, OzakiKino2005}. Exchange and 
correlation effects were treated within the Perdew-Burke-Ernzerhof 
generalized gradient approximation 
(PBE-GGA)~\cite{pbe}. The supercell used for each 
slab has the form $3 \times N \times 3$, where $N$ is the number of 
layers composing the slab. A Monkhorst-Pack 
mesh~\cite{monkhorst1976special} of $3 \times 1 \times 3$ was used 
to sample the Brillouin zone, with a single $k$-point along the 
non-periodic slab direction. Tellurium atoms were described using a 
fully relativistic norm-conserving pseudopotential and an optimized 
pseudo-atomic basis set with a confinement radius of $7.0$~Bohr, 
comprising three $s$, three $p$, two $d$, and one $f$ radial 
functions per atom, both taken from the OpenMX 2019 
database~\cite{Ozaki2003, OzakiKino2004}. Calculations were performed 
in a non-spin-polarized, scalar-relativistic setting, since the 
present analysis focuses on charge redistribution rather than on 
spin- or spin-orbit-induced features. The real-space cutoff energy 
was set to $150$~Ha ($300$~Ry), corresponding to a grid spacing of 
approximately $0.10$~\AA. This grid was used to represent both the 
Hartree potential associated with the difference charge density, 
$\delta V_{\mathrm{H}}(\mathbf{r})$, and the difference charge 
density $\delta\rho(\mathbf{r})$ itself, which serve as inputs to the 
local field analysis described below.

\subsection{Local Electric Field at Atomic Sites}

The vectors $\mathbf{E}_i$ shown at each atomic site in 
Fig. 3 were computed in three steps from the 
self-consistent Hartree potential of the difference charge density, 
$\delta V_{\mathrm{H}}$, written by OpenMX in \texttt{.cube} format. 
We use $\delta V_{\mathrm{H}}$ rather than the full Hartree potential 
because the spherical neutral-atom contribution produces no net field 
at the sites~\cite{OzakiTotalEnergy}, so
\begin{equation}
\mathbf{E}(\mathbf{r}) = -\nabla\,\delta V_{\mathrm{H}}(\mathbf{r})
\label{eq:Efield}
\end{equation}
isolates the field associated with bonding-induced charge 
redistribution.

First, we differentiate the potential on the \texttt{.cube} grid, 
which is regular only in fractional coordinates. We use a hybrid 
scheme: spectral (FFT) differentiation along the two periodic axes, 
exploiting
\begin{equation}
\partial_\alpha \tilde V \;\longrightarrow\; 
(2\pi i\,k_\alpha)\,\hat{\tilde V},
\label{eq:fftdiff}
\end{equation}
and a centered sixth-order finite-difference stencil along the slab 
direction, with one-sided formulas near the vacuum boundaries.

Second, we evaluate the gradient at each atomic position $\mathbf{p}_i$ 
without interpolating the precomputed grid. A two-dimensional FFT of 
$\tilde V$ over the periodic axes yields the corresponding derivatives 
along the entire slab-direction column through the atom,
\begin{equation}
\partial_\alpha \tilde V_k(\mathbf{u}_0) = \frac{1}{n_x n_z}
\sum_{\mathbf{q}}(2\pi i\,k_\alpha)\,\hat V_{\mathbf{q},k}\,
e^{2\pi i\,\mathbf{q}\cdot\mathbf{u}_0},
\label{eq:synthesis}
\end{equation}
where $\mathbf{u}_0 = (u_0, w_0)$ is the in-plane projection of the 
atomic position and $\mathbf{q} = (p,q)$ runs over the in-plane 
Fourier modes, to machine precision; a quintic spline along that 
column then provides the remaining derivative at the atomic ordinate, 
with residual error $\mathcal{O}(h^5)$.

Third, we convert the fractional gradient to Cartesian components 
through
\begin{equation}
\nabla_{\mathbf{r}}\,\delta V_{\mathrm{H}} = M^{-\mathsf{T}}\,
\nabla_{\mathbf{s}}\tilde V,
\label{eq:Minvtrans}
\end{equation}
where $M$ is the cell matrix. The off-diagonal entries of 
$M^{-\mathsf{T}}$ are essential for the oblique helical-Te cell; 
omitting them produces errors of order unity in the $y$ component. 
Applying the Hartree-to-eV conversion then yields $\mathbf{E}_i$ in 
V/\AA. The result was cross-checked against an independent tricubic 
interpolation of the precomputed gradient and against the residual 
$C_3$ symmetry of the helical axis.

\begin{figure}[t]
    \includegraphics[width=0.93\linewidth]{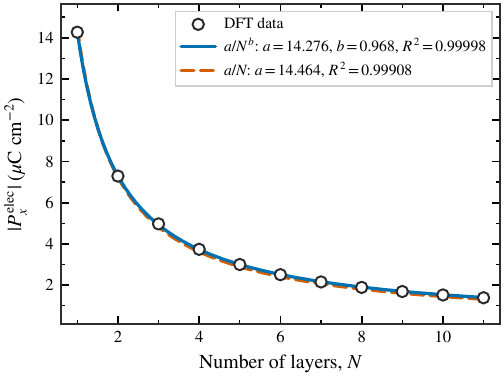}
    
    \makeatletter
    \renewcommand{\fnum@figure}{\figurename~S\thefigure}
    
    \caption{Magnitude of the electronic polarization component along 
    $x$, $|P_x^{\mathrm{elec}}|$, as a function of the number of 
    layers $N$ composing the slab. Open circles are DFT results; the 
    solid (blue) and dashed (orange) curves are fits to $a/N^b$ and 
    $a/N$, respectively, with the fitted parameters indicated in the 
    legend.}
    \label{fig:SI:polxN}
\end{figure}

\subsection{Polarization as a Function of the Number of Layers}
The macroscopic polarization along the $x$ direction, perpendicular 
to the Te helices, was computed within the modern theory of 
polarization~\cite{KingSmithVanderbilt1993,Resta1994}, as implemented 
in OpenMX~\cite{OzakiBerryPhase}. The electronic contribution is 
given by the Berry phase
\begin{equation}
\mathbf{G}_i \cdot \mathbf{P}^{\mathrm{el}} = 
-\frac{e}{(2\pi)^3}\sum_\sigma \int_{\mathrm{BZ}} d^3k\;
\mathbf{G}_i \cdot \left(\frac{\partial}{\partial \mathbf{k}'}
\eta_\sigma(\mathbf{k},\mathbf{k}')\right)_{\mathbf{k}'=\mathbf{k}},
\label{eq:berry}
\end{equation}
where $\mathbf{G}_i$ are the reciprocal-lattice vectors and 
$\eta_\sigma(\mathbf{k},\mathbf{k}') = \mathrm{Im}\,\ln\det\langle 
u^{(\mathbf{k})}_{\sigma\mu}|u^{(\mathbf{k}')}_{\sigma\nu}\rangle$ is 
the gauge-invariant phase of the overlap matrix between the 
cell-periodic parts of the occupied Bloch states. The Brillouin-zone 
integral was evaluated through the discretized 
King-Smith--Vanderbilt formula on a string of $N_1$ $k$-points along 
$\mathbf{G}_1$, sampled at $N_2 \times N_3$ points in the transverse 
plane.

We repeated this calculation for slabs of $N = 1, 2, \ldots, 11$ 
layers, keeping the in-plane lattice parameter, vacuum thickness, and 
self-consistency convergence criteria identical to those of the 
single-layer calculation. The resulting electronic polarization 
component $|P_x^{\mathrm{elec}}(N)|$ is shown in 
Fig. S\ref{fig:SI:polxN}. The data are well described by an 
inverse-layer dependence, $|P_x^{\mathrm{elec}}| \approx a/N$ with 
$a \approx 14.5~\mu\mathrm{C\,cm^{-2}}$ ($R^2 = 0.999$); a 
free-exponent fit yields $b = 0.97$, confirming that the deviation 
from a pure $1/N$ law is negligible. This $1/N$ scaling is the slab 
analogue of the $\propto 1/t$ thickness dependence observed 
experimentally in the main text.

\end{document}